\documentclass[12pt]{extarticle}   
\usepackage[margin=1in]{geometry}

\usepackage{amssymb}
\usepackage{amsmath}
\usepackage{multirow}
\usepackage{latexsym}
\usepackage{epsfig}
\usepackage{graphicx}
\usepackage{epstopdf}
\usepackage[usenames]{color}

\usepackage{array}
\usepackage{natbib}
\usepackage{amsthm}
\usepackage{rotating}
\usepackage{multirow}
\usepackage{bbold}
\usepackage{verbatim}
\usepackage{enumitem}  
\usepackage{outlines}
\usepackage[colorlinks=true,linkcolor=blue,citecolor=blue]{hyperref}	
\usepackage{changepage}  
\usepackage{longtable}  
\usepackage{pdflscape}  
\usepackage{multicol}  
\usepackage{xtab}
\usepackage{algorithm}
\usepackage{algorithmic}
\usepackage{physics}   

\usepackage{booktabs}
\usepackage{tabularx}
\usepackage{threeparttable}
\usepackage{adjustbox}

\usepackage{amsmath}

\newcommand{\bfalpha}{\mbox{\boldmath $\alpha$}}		
\newcommand{\bfbeta}{\mbox{\boldmath $\beta$}}      	
\newcommand{\bfgamma}{\mbox{\boldmath $\gamma$}}	
\newcommand{\bfdelta}{\mbox{\boldmath $\delta$}}

\newcommand{\bfeta}{\mbox{\boldmath $\eta$}}		
\newcommand{\bftheta}{\mbox{\boldmath $\theta$}}		
		
\newcommand{\bfkappa}{\mbox{\boldmath $\kappa$}}

\newcommand{\bfupsilon}{\mbox{\boldmath $\upsilon$}}

\newcommand{\bfpsi}{\mbox{\boldmath $\psi$}}

\newcommand{\bfvartheta}{\mbox{\boldmath $\vartheta$}}

			\newcommand{\bfB}{{\bf B}}
			\newcommand{\bfD}{{\bf D}}

	\newcommand{\bfM}{{\bf M}}		
			
	\newcommand{\bfQ}{{\bf Q}}			
		\newcommand{\bft}{{\bf t}}		
	\newcommand{\bfU}{{\bf U}}		
\newcommand{\bfw}{{\bf w}}		\newcommand{\bfx}{{\bf x}}	\newcommand{\bfX}{{\bf X}}
\newcommand{\bfy}{{\bf y}}		\newcommand{\bfz}{{\bf z}}	\newcommand{\bfZ}{{\bf Z}}

\allowdisplaybreaks   
\makeatletter
\let\mcnewpage\newpage
\newcommand{\changenewpage}{%
  \renewcommand\newpage{%
    \if@firstcolumn
      \hrule width\linewidth height0pt
      \columnbreak
    \else
      \mcnewpage
    \fi
}}
\makeatother

\usepackage{hyperref} 

\begin{document}

\clearpage

\begin{center}
{\Large \bf Competing-Risk Cure Models: A Comprehensive\\[4pt] Systematic Review of Methodological Literature}  

Nilotpal Sanyal\\
Department of Mathematical Sciences,\\
University of Texas at El Paso, Texas, USA\\
nsanyal@utep.edu
\end{center}

\begin{abstract}
Competing-risk cure models describe time-to-event populations in which some individuals are not susceptible to all event types or to a particular event of interest, but the methodological literature has developed through largely separate model families. We review 26 methodological contributions organized along five axes---the definition and scope of cure; the decomposition of the joint distribution and cure mechanism; the latency structure; the treatment of dependence, censoring, and masked causes; and the estimation framework. The review distinguishes global from cause-specific cure and separates incidence--latency mixtures from vertical susceptibility factorizations, latent competing-causes or zero-count constructions, defective-survival models, and zero-inflated mixture or cumulative incidence function (CIF) formulations. It compares parametric, piecewise-constant, PH, AFT, transformation, CIF-based, nonparametric, and partially specified latency models, including approaches for right- or interval-censored, clustered, and masked-cause data.

Mixture formulations remain dominant, but similarly named models may target different estimands and entail distinct cure mechanisms, latent-risk and censoring assumptions, and regression interpretations. Dependence among latent failure times is less often modeled explicitly than cure or latency, whereas failure--censoring dependence, within-cluster association, and masked causes receive targeted treatment in smaller subsets of models. Estimation ranges from likelihood and expectation-maximization (EM) methods, including neural-network M-steps, to estimating equations, inverse-probability-of-censoring weighting, Bayesian computation, and copula-graphic estimation. A reproducible defective-Gompertz illustration using public bone-marrow-transplant data shows why fitted tail probabilities must be interpreted according to the model and endpoint definition rather than as an all-method comparison. Reproducibility remains limited---most implementations rely on custom code, few provide repository-level access, and no widely adopted, clearly licensed R or Python framework unifies the principal constructions. By organizing the literature, this taxonomy helps researchers select and report models transparently, compare estimation approaches, and identify needs for further theory, software, benchmarking, and reproducible applications.
\end{abstract}

\noindent\textit{Keywords:} competing risks; cure models; long-term survival; dependent censoring; mixture cure models; incidence-latency mixture; latent competing causes; defective survival models.


%
%
%
\setlength{\parskip}{2pt}
%
%
%


\section{Introduction}
Time-to-event methods are used to study the occurrence and timing of events such as death, relapse, disease progression, or system failure. In many applications, however, an individual may experience one of several mutually exclusive event types, called competing risks scenarios, where the occurrence of one event precludes the occurrence or observation of the others. The observed outcome is determined by both the event-time distribution and the mechanism assigning the event type. Consequently, the probability of experiencing a particular event by a certain time is a cumulative incidence function (CIF), not simply a survival probability obtained by treating other causes as censoring \citep{kalbfleisch_prentice_2002, putter_et_al_2007}. For example, treating competing events as independent censoring in a Kaplan--Meier analysis can overestimate the incidence of the event of interest because it removes individuals who can no longer experience that event.

A second complication arises when a subset of individuals is not susceptible to the event under study. Such individuals are often described as cured or as long-term survivors, a concept underlying the classical mixture-cure literature \citep{boag_1949, berkson_gage_1952, farewell_1982, maller_zhou_1996, sy_taylor_2000}. The term ``cure'' is model-dependent---it may mean immunity to all competing causes, immunity only to a primary cause while other causes remain possible, or the absence of all latent failure-initiating causes. Thus, a plateau in an empirical survival curve cannot by itself establish cure. A cure interpretation requires a statistical model, tail assumptions, and sufficient follow-up or other identifying information. Classical mixture cure models represent cure through an incidence or susceptibility component and model latency among susceptible individuals, whereas other constructions can encode cure through structural susceptibility, a zero latent count, a defective survival distribution, or an explicit zero-inflated mass.

The simultaneous presence of competing risks and cure fractions creates a more fundamental modeling problem than either feature considered alone. A model must specify how cure status, event time, and event type are related, whether cure is global or cause-specific, whether it introduces a joint structure for latent risks, and how censoring, within-cluster association, and masked causes are handled. These choices affect the interpretation of the cure fraction, CIF, overall survival function, and regression parameters. They also affect identifiability, because only the minimum of any latent event times and the cause of the observed event are typically available.

Early competing-risk cure models extended the classical mixture-cure idea by allowing either a global cured class or cure from a primary event while a competing event remained possible. Subsequent work broadened the literature to semiparametric and Bayesian formulations, direct-CIF and vertical models, latent-count and defective-survival constructions, masked causes, dependent censoring, clustering, interval censoring, and zero inflation. These developments show why similarly named models can differ in cure target, joint-distribution construction, event-time structure, and observation assumptions.

Although this literature has produced substantial theoretical and methodological advances, it remains fragmented. The same broad label---for example, ``mixture cure model''---can refer to materially different definitions of cure, decompositions of the joint distribution, latent-risk or censoring assumptions, latency structures, and fitting procedures. Conversely, models not labeled as mixture models may produce long-term survival through defective distributions or latent counts, and zero-inflated mixture or CIF models combine still other ingredients. Comparisons based only on a latency distribution or cure link therefore do not fully reveal the similarities and differences among methods. Reproducibility is also uneven---most implementations rely on custom code, and publicly available, clearly licensed software providing a unified competing-risk cure framework remains limited.

This methodological review synthesizes 26 contributions, from early mixture formulations to recent semiparametric, Bayesian, defective, latent-count, zero-inflated, clustered, interval-censored, and neural models. Its central objective is to provide a common conceptual and mathematical framework for comparing these methods. We organize the literature along five complementary axes:

\begin{enumerate}
\item the definition and scope of cure, including global versus cause-specific cure and the treatment of noncurable competing risks;
\item the decomposition of the joint distribution of event time, event type, and cure status, including mixture, zero-inflated mixture/CIF, vertical, latent competing-causes, and defective-survival constructions;
\item the structure of the event-time (latency) component, including parametric and piecewise-constant, PH, AFT, transformation, CIF-based, nonparametric, and partially specified forms;
\item dependence and observation-process assumptions, including dependence among latent risks, associations between cure status and the event process, the relation between failure and censoring, within-cluster dependence, and the treatment of masked causes; and
\item the estimation framework, including direct and partial likelihood, semiparametric and nonparametric likelihood, EM (including neural M-steps), estimating equations/IPCW, Bayesian computation, and copula-graphic estimation.
\end{enumerate}

Table~\ref{tab:five_axes_classification} summarizes the classification of the 26 reviewed contributions along these five axes. For each axis, we distinguish assumptions that are often conflated in the literature. In particular, we separate a model for latent competing-risk dependence from a model for failure--censoring dependence, distinguish an explicit cure indicator from a defective-distribution tail mass and a zero-count probability, and distinguish EM point estimation from Bayesian data augmentation targeting a posterior distribution. We also review the computational tools and code-availability statements reported by the individual papers to assess the current state of reproducibility.

We also include a reproducible illustration using an explicitly constrained defective Gompertz competing-risk cure construction motivated by \cite{silpa_et_al_2024} and a public bone-marrow-transplant data set. The method was selected because author-provided R code can be directly adapted, and the illustration demonstrates the five axes rather than attempting an all-method comparison on one data set.

The remainder of the manuscript develops these comparisons in detail. The first axis establishes the meaning and scope of cure. The second examines decompositions of the joint distribution and the mechanisms that generate cure. The third compares latency structures, the fourth examines dependence, censoring, and masked causes, and the fifth compares estimation frameworks and uncertainty quantification. The manuscript also presents the real-data illustration and a software and reproducibility assessment that evaluates the availability and verifiability of implementations and identifies priorities for a unified computational framework.

\begin{landscape}
\begin{table}
\centering
\caption{Classification of Competing Risk Cure Models Along Five Fundamental Axes}
\begin{adjustbox}{max width=1.5\textwidth}
\begin{threeparttable}
\begin{tabular}{lccccc}
\toprule
\textbf{Paper} 
& \textbf{Axis 1} & \textbf{Axis 2} & \textbf{Axis 3} & \textbf{Axis 4} & \textbf{Axis 5} \\
\midrule
\cite{greenhouse_wolfe_1984} & Cause-specific cure & Mixture & Parametric (Weibull) & Conditional latent-risk indep. & Direct MLE \\
\cite{chao_1998} & Global cure & Mixture & Parametric (exponential) & No cross-cause joint model & Bayesian MCMC \\
\cite{ng_mclachlan_1998} & Global cure & Mixture & Partially specified & No cross-cause joint model & Partial likelihood \\
\cite{choi_zhou_2002} & Global cure & Mixture & Parametric & No cross-cause joint model & Direct MLE \\
\cite{li_et_al_2007} & Global cure & Mixture & Nonparametric & Failure--censoring copula & Copula-graphic estimation \\
\cite{othus_et_al_2009} & Global cure & Mixture & Transformation (semi-param.) & Dependent censoring (IPCW) & IPCW estimating eqs. \\
\cite{basu_tiwari_2010} & Cause-specific cure & Mixture & Parametric / piecewise constant & Masked causes; cure--event link & Bayesian MCMC \\
\cite{choi_et_al_2015} & Global cure & Mixture & Transformation (semi-param.) & No cross-cause joint model & Semiparametric NPMLE \\
\cite{choi_et_al_2018} & Global cure & Mixture & AFT (semi-param.) & No cross-cause joint model & Kernel-smoothed profile likelihood \\
\cite{nicolaie_et_al_2019} & Global cure & Vertical & PH total hazard & No baseline cross-cause joint model & EM \\
\cite{wang_et_al_2020a} & Cause-specific cure & Mixture & PH & No cross-cause joint model & Bayesian MCMC / augmentation \\
\cite{wang_et_al_2020b} & Cause-specific cure & Mixture & AFT (semi-param.) & No cross-cause model; conditional censoring & EM \\
\cite{rejani_sankaran_2020} & Global cure & Mixture & PH & No cross-cause joint model & EM \\
\cite{kuttumannil_et_al_2020} & Global cure & CIF-based transformation & CIF transformation & No cross-cause joint model & Counting-process estimating eqs. \\
\cite{chen_et_al_2020} & Global cure & Mixture / selected CIF & CIF-based (semi-param.) & Dependent censoring (IPCW) & Two-stage IPCW estimating eqs. \\
\cite{wang_et_al_2021} & Cause-specific cure & Mixture & PH & No cross-cause joint model & EM \\
\cite{esmailian_et_al_2023} & Cause-specific cure & Latent competing causes & Parametric (GeTNH) & Conditional i.i.d. latent causes & Direct MLE \\
\cite{menger_et_al_2023} & Cause-specific cure & Cause-specific mixture & PH (piecewise baseline) & Masked causes; cure--event link & Bayesian MCMC / augmentation \\
\cite{silpa_et_al_2024} & Global cure & Defective survival & Parametric (defective) & Independent cause components & Direct MLE (BFGS) \\
\cite{silpa_et_al_2024a} & Global cure & Defective survival & Parametric (defective) & Independent cause components & Direct MLE (\texttt{nlminb}) \\
\cite{teh_et_al_2025} & Global cure & Latent competing causes & Piecewise exponential & Conditional i.i.d. latent causes & EM with neural-network M-step \\
\cite{ganguly_et_al_2026} & Global cure & Finite mixture & Parametric (Weibull mixture) & Mode-mixture dependence; noninformative censoring & EM \\
\cite{pal_roy_2026} & Global cure & Latent causes (destructive) & Parametric (Weibull) & Independent latent progressions; random censoring & Direct likelihood (SQH) \\
\cite{sreedevi_et_al_2026} & Global cure + zero inflation & Zero-inflated mixture/CIF & Parametric (Gompertz CIF) & Independent causes/censoring & Constrained MLE (\texttt{optim}) \\
\cite{wang_et_al_2026a} & Cause-specific cure & Clustered mixture & PH (semi-param.) & Working cluster correlation & ES/GEE + bootstrap \\
\cite{wang_et_al_2026b} & Global cure & Multinomial mixture & Parametric AFT & No cross-cause model; conditional censoring & EM \\
\bottomrule
\end{tabular}
{\footnotesize\textit{Note.} Axis~1: definition of cure; Axis~2: decomposition of the joint distribution and cure modeling; Axis~3: latency model structure; Axis~4: treatment of dependence, censoring, and masked causes; and Axis~5: estimation framework. ``No cross-cause joint model'' indicates that the model specifies the observed event-time/cause distribution without an association model for counterfactual latent failure times.}
\end{threeparttable}
\end{adjustbox}
\label{tab:five_axes_classification}
\end{table}
\end{landscape}

\section{Axis 1: Definition of Cure}
\label{sec:axis1}

A first and foundational axis in competing-risk cure modeling concerns what it means for an individual to be cured when multiple mutually exclusive event types are possible. Across the 26 reviewed contributions, two principal definitions emerge: \emph{global cure}, in which cured individuals are immune to all competing event types, and \emph{cause-specific cure}, in which cure applies only to a specified primary event, often called the event of interest, while other competing events remain possible.

The global-cure interpretation is used by \citep{chao_1998, ng_mclachlan_1998, choi_zhou_2002, li_et_al_2007, othus_et_al_2009, choi_et_al_2015, choi_et_al_2018, nicolaie_et_al_2019, chen_et_al_2020, kuttumannil_et_al_2020, rejani_sankaran_2020, silpa_et_al_2024, silpa_et_al_2024a, teh_et_al_2025, ganguly_et_al_2026, pal_roy_2026, sreedevi_et_al_2026, wang_et_al_2026b}. For these models, the cure probability is the probability of remaining free from all modeled events indefinitely. Axis~2 classifies the distinct mechanisms through which this global-cure probability is generated.

The cause-specific-cure interpretation is used by \citep{greenhouse_wolfe_1984, basu_tiwari_2010, wang_et_al_2020a, wang_et_al_2020b, wang_et_al_2021, esmailian_et_al_2023, menger_et_al_2023, wang_et_al_2026a}. A person cured of the primary event may therefore still be vulnerable to an incurable competing event. For example, in \cite{greenhouse_wolfe_1984}, cure applies to disease death but not to death from other causes. \cite{basu_tiwari_2010} likewise define cure with respect to the primary cause while allowing distinct subhazards for the competing causes and cure groups. \citep{wang_et_al_2020a, wang_et_al_2020b, wang_et_al_2021} formulations allow a primary event with a potential cure fraction together with a noncurable competing event; in the clustered marginal model of \cite{wang_et_al_2026a}, the cluster-specific latent indicator $Y_{ij}$ identifies susceptibility to the primary event. Some models allow a separate cure status for every cause, as in \cite{menger_et_al_2023}. The remaining axes classify, separately, how the chosen cure definition is embedded in the joint distribution, the latency model, the dependence assumptions, and the estimation framework.

\section{Axis 2: Decomposition of the Joint Distribution and Cure Modeling}
A second fundamental axis along which competing-risk cure models differ is the decomposition of the joint distribution of event times and event types, and the consequent mechanism for modeling cure. This axis concerns how the cure phenomenon is encoded mathematically within the survival framework and how it is linked to the competing-risk process. In what follows, vectors and matrices are denoted by boldface letters, whereas scalar quantities are denoted by ordinary italic letters.

Let $T$ denote the time to the first observed event, $E\in\{1,\ldots,K\}$ the event type, $C$ a censoring time, and $\delta=I(T\le C)$. Let $Z\in\{0,1\}$ denote a latent cure indicator, where $Z=1$ represents cure under the model-specific definition adopted in Axis~1, and let $\bfX$ denote the covariate vector. The modeling problem centers on how the joint distribution
$$
P(T,E,Z\mid \bfX)
$$
is factorized, parameterized, and linked to regression components. Across the 26 reviewed contributions, four principal decomposition strategies emerge: \emph{incidence--latency (mixture) decomposition}, \emph{vertical (susceptibility) decomposition}, \emph{latent competing-causes (promotion-time) construction}, and \emph{defective (improper) survival decomposition}. Some papers combine or approximate more than one of these ideas, but they differ conceptually in how cure interacts with the competing-risk mechanism.

\subsection{Incidence--Latency (Mixture) Decomposition}
The dominant approach in the literature adopts a mixture-based incidence--latency decomposition \citep{greenhouse_wolfe_1984, chao_1998, ng_mclachlan_1998, choi_zhou_2002, li_et_al_2007, othus_et_al_2009, basu_tiwari_2010, choi_et_al_2015, choi_et_al_2018, wang_et_al_2020a, wang_et_al_2020b, wang_et_al_2021, rejani_sankaran_2020, chen_et_al_2020, menger_et_al_2023, sreedevi_et_al_2026}. The part that determines whether an event can occur is called the incidence part, and the part that determines the time of occurrence of an event, given that an event can occur, is called the latency part \citep{sy_taylor_2000}. In its most general form, the decomposition is a weighted combination of a cured group and a susceptible group,

\begin{equation}
P(T,E\mid \bfX)=\pi(\bfX)P(T,E\mid Z=1,\bfX)
 +\{1-\pi(\bfX)\}P(T,E\mid Z=0,\bfX),
\label{eq:joint_mixture}
\end{equation}

where $\pi(\bfX)=P(Z=1\mid \bfX)$ denotes the cure probability, or incidence component, and the second term specifies the latency distribution among the uncured individuals ($Z=0$). In survival-function form,

\begin{equation}
S(t\mid \bfX)=\pi(\bfX)+\{1-\pi(\bfX)\}S_u(t\mid \bfX),
\label{eq:mixture}
\end{equation}

where $S_u(t\mid \bfX)$ is the survival function among susceptible (uncured) individuals.

The interpretation of $\pi$ follows Axis~1. Under global cure, the cured component places all probability mass at $T=\infty$. Under cause-specific cure, it excludes only the event of interest ($E=1$) while retaining the other risks, so that, for example,

\begin{equation}
P(T<\infty,E=1\mid Z=1,\bfX)=0,
\label{eq:cause_specific_mixture}
\end{equation}

while other causes may remain possible. 

Among the studies using the above mixture decomposition, \cite{greenhouse_wolfe_1984, chao_1998, ng_mclachlan_1998, li_et_al_2007, basu_tiwari_2010} do not specify the cure or incidence component as a covariate-based regression. In \cite{greenhouse_wolfe_1984}, the cure probability $p$ is assumed constant, although the authors note that a more general formulation could allow it to depend on age. Similarly, \cite{ng_mclachlan_1998} mention a possible logistic covariate extension for $p$ but neither specify nor fit such a model. The remaining three papers do not formulate the cure or incidence component as a regression on covariates. 

For the studies that do use regression-based cure or incidence specifications, the corresponding formulations are given below.
 

\paragraph{Baseline-Category Multinomial Logit.}
The earliest specification in this group is that of \cite{choi_zhou_2002}. With $B_i=0$ denoting immunity from all $J$ causes, $B_i=j$ denoting eventual failure from cause $j$, and $P(B_i=j)=p_{ij}$, they specify

\begin{equation}
p_{ij} = \frac{\exp(\bfbeta_j^\top \bfy_i)}{1+\sum_{\ell=1}^{J}\exp(\bfbeta_\ell^\top \bfy_i)},\qquad i=1,\ldots,n,\quad j=1,\ldots,J,
\end{equation}

where $\bfy_i$ is the $k_2$-vector of covariates and $\bfbeta_j$ is the corresponding $k_2$-vector of regression coefficients. The cure probability is therefore the baseline-category probability

\begin{equation}
P(B_i=0)=1-\sum_{j=1}^{J}p_{ij}=\frac{1}{1+\sum_{j=1}^{J}\exp(\bfbeta_j^\top \bfy_i)}.
\end{equation}

\cite{choi_et_al_2015}, \cite{choi_et_al_2018}, \cite{rejani_sankaran_2020}, and \cite{wang_et_al_2026b} use the same baseline-category multinomial-logit incidence structure.
In each case, cure is the baseline class and the noncure event classes receive the multinomial-logit probabilities.

\paragraph{Binary Logistic Susceptibility Model.}
\cite{othus_et_al_2009} use $\eta=1$ for an uncured subject and specify the susceptibility probability as

\begin{equation}
P(\eta=1\mid \bfX,\bfZ)=G(\bfgamma^\top \bfX),\qquad G(u)=\frac{\exp(u)}{1+\exp(u)},
\end{equation}

where $\bfX$ is the vector of covariates associated with cure indicator $\eta$ and $\bfZ$ is the vector of covariates associated with survival time.
Thus their cure probability is $P(\eta=0\mid \bfX,\bfZ)=1-G(\bfgamma^\top \bfX)$. \cite{chen_et_al_2020} and \cite{ganguly_et_al_2026} use the same binary-logistic incidence specification, where the latter models the cure probability directly.

\paragraph{Primary-Event Susceptibility Plus Conditional Event-Class Logit.}
For a primary event that is curable and a competing event that is not, \cite{wang_et_al_2020b} use two logistic regressions. Using a latent indicator $Y_i=1$ to denote that subject $i$ is uncured with respect to the primary event, they specify

\begin{equation}
\pi_\alpha(\bfX_i)\equiv P(Y_i=1\mid \bfX_i) = \frac{\exp(\bfalpha^\top\widetilde \bfX_i)}{1+\exp(\bfalpha^\top\widetilde \bfX_i)},\qquad \widetilde \bfX_i=(1,\bfX_i^\top)^\top.
\end{equation}

Conditional on $Y_i=1$, they then specify the event class by

\begin{equation}
\pi_\theta(\bfX_i)\equiv P(\epsilon_i=1\mid \bfX_i,Y_i=1) = \frac{\exp(\bftheta^\top\widetilde \bfX_i)}{1+\exp(\bftheta^\top\widetilde \bfX_i)}.
\end{equation}

Accordingly, the primary-event cure probability is $1-\pi_\alpha(\bfX_i)$, while the marginal event-class probabilities are

\begin{equation}
P(\epsilon_i=1\mid \bfX_i) = \pi_\alpha(\bfX_i)\pi_\theta(\bfX_i),\qquad P(\epsilon_i=2\mid \bfX_i) = 1-\pi_\alpha(\bfX_i)\pi_\theta(\bfX_i).
\end{equation}

\cite{wang_et_al_2020a} and \cite{wang_et_al_2021} retain this same two-logit incidence construction.

\paragraph{Zero-Inflated Mixture/CIF Component Allocation.}
\cite{sreedevi_et_al_2026} should be kept separate from the ordinary incidence--latency mixture specifications because their model combines zero-inflation through cause-specific point masses at time zero, a continuous susceptible component, and a positive tail mass for global cure, thereby distinguishing immediate failures, long-term survivors, and susceptible subjects who fail later. With $p_{0j}$ denoting the probability of an immediate failure from cause $j$, $p_1$ the global cure probability, and $D(\bfx)=1+\sum_{\ell=1}^{K}\exp(\bfx^\top\bfeta_\ell)+\exp(\bfx^\top\bfbeta)$, they specify

\begin{equation}
p_{0j}=\frac{\exp(\bfx^\top\bfeta_j)}{D(\bfx)},\qquad p_1=\frac{\exp(\bfx^\top\bfbeta)}{D(\bfx)},\qquad j=1,\ldots,K.
\end{equation}

Here $\bfeta_j$ and $\bfbeta$ are the regression coefficients for the cause-$j$ zero-inflation probability and the cure probability, respectively. The total immediate-failure probability is $p_0=\sum_{j=1}^{K}p_{0j}$, and the residual probability $1-p_0-p_1$ is assigned to the continuous susceptible component. Algebraically, this is a baseline-category multinomial logit with continuous susceptibility as the reference category. It is presented separately so that its zero-inflation mechanism is not conflated with an ordinary event-class incidence model. Its population survival can equivalently be written as

\begin{equation}
S(t\mid \bfX)=p_1(\bfX)+\{1-p_0(\bfX)-p_1(\bfX)\}S^*(t\mid \bfX),\qquad p_0(\bfX)=\sum_{j=1}^K p_{0j}(\bfX).
\label{eq:zero_inflated_cure_contrast}
\end{equation}

Thus, the positive tail is the explicit global-cure mass $p_1(\bfX)$, not a tail induced by an improper baseline distribution.

\paragraph{Separate Cause-Specific Cure Logits.}
\cite{menger_et_al_2023} specify a distinct cure probability for every cause. With $d_k=1$ indicating cure from cause $k$ and $\bfx_0$ denoting baseline covariates including an intercept, their cure-rate submodel is

\begin{equation}
\pi_k(\bfx_0\mid\bfbeta_k^C) = P(d_k=1\mid \bfx_0,\bfbeta_k^C) = \frac{\exp(\bfx_0^\top\bfbeta_k^C)}{1+\exp(\bfx_0^\top\bfbeta_k^C)},\qquad k=1,\ldots,K.
\end{equation}

The model's overall cure probability is $\prod_{k=1}^{K}\pi_k(\bfx_0\mid\bfbeta_k^C)$, whereas $\pi_k(\bfx_0\mid\bfbeta_k^C)$ itself is the cause-specific cure probability.

\subsection{Vertical (Susceptibility) Factorization}

A second decomposition emphasizes a susceptibility layer above the competing-risk mechanism. This factorization is central to the vertical modeling framework of \cite{nicolaie_et_al_2019}. Its distinctive feature is the conditional factorization

\begin{equation}
P(T,E\mid Y=1,\bfX)=P(E\mid T,Y=1,\bfX)P(T\mid Y=1,\bfX),
\label{eq:vertical_event_factorization}
\end{equation}

rather than the conventional competing-risk mixture decomposition $P(T,E)=P(T\mid E)P(E)$, which conditions on a future event type presumed to be determined at baseline. Under the vertical perspective, susceptibility governs entry into the competing-risk hazard system, and the event type is assigned only when a failure occurs. This avoids imposing a baseline joint distribution for future cause labels and is the defining contribution of the vertical approach.

The resulting joint distribution can be written schematically as

\begin{equation}
P(T,E\mid \bfX)=P(Y=0\mid \bfX)\,\delta_{\infty}(T)+P(Y=1\mid \bfX)P(E\mid T,Y=1,\bfX)f(T\mid Y=1,\bfX),
\label{eq:vertical_factorization}
\end{equation}

where $Y=0$ denotes cure, $f(T\mid Y=1,\bfX)$ is the density of the event time among susceptible individuals, and $\delta_{\infty}$ denotes a point mass at infinity. Rather than representing cure solely through an additive survival mass, this formulation conceptualizes cure as a structural susceptibility indicator, where susceptibility acts multiplicatively on the total hazard among the at-risk population. Although mathematically related to the mixture representation, the particular regression structures for the susceptible event-time and relative-cause components are classified under Axis~3.

\subsection{Latent Competing-Causes (Promotion-Time) Construction}
A third strategy encodes cure through a latent number of failure-initiating causes. Here, competing causes are unobserved risks that can initiate the event, rather than necessarily the observed event types $E$ defined above. Within the reviewed literature, \cite{esmailian_et_al_2023, pal_roy_2026, teh_et_al_2025} each define cure through a zero latent count, but they implement this principle differently---\cite{esmailian_et_al_2023} mixes a Poisson count, \cite{pal_roy_2026} thins an initial count after treatment, and \cite{teh_et_al_2025} uses Poisson and negative-binomial count models. In this section, $S$ denotes the relevant proper latent progression-time survival function and $F=1-S$ its distribution function.

\paragraph{Poisson--Gamma and Poisson--generalized-exponential construction.} 
In \cite{esmailian_et_al_2023}, $M$ is the unobserved number of competing causes, $R_j$, $j=1,\ldots,M$, are independent latent event times with survival function $S(t)$, and $M=0$ denotes cure. Equivalently, $P(R_0=\infty)=1$ and $T=\min\{R_0,R_1,\ldots,R_M\}$. Their mixed-Poisson formulation is

\begin{equation}
M\mid\eta\sim\operatorname{Poisson}(\eta\theta),\qquad G_M(s)=\mathcal{M}_{\eta}\{-\theta(1-s)\},\qquad S_{\mathrm{pop}}(t)=G_M\{S(t)\},
\label{eq:promotion_time}
\end{equation}

where $G_M$ is the marginal probability-generating function, $\mathcal{M}_{\eta}$ is the moment-generating function of the unobserved heterogeneity variable $\eta$, and $S_{\mathrm{pop}}$ is the population survival function. Thus,

\begin{equation}
p_0=P(M=0)=G_M(0)=\mathcal{M}_{\eta}(-\theta).
\label{eq:promotion_cure}
\end{equation}

For their Poisson--Gamma cure-rate model, with $\eta \sim \text{Gamma}(\sigma,\sigma)$, the Gamma mixing distribution gives

\begin{equation}
S_{\mathrm{pop,PG}}(t)=\{1+(\theta/\sigma)F(t)\}^{-\sigma},\qquad p_{0,\mathrm{PG}}=\{1+\theta/\sigma\}^{-\sigma},
\label{eq:promotion_pg}
\end{equation}

whereas their Poisson--generalized-exponential cure-rate model, with $\eta$ following generalized-exponential, gives

\begin{equation}
S_{\mathrm{pop,PGE}}(t)=\frac{2}{[2+(\theta/\gamma)F(t)][1+(\theta/\gamma)F(t)]},\qquad p_{0,\mathrm{PGE}}=\frac{2}{(2+\theta/\gamma)(1+\theta/\gamma)}.
\label{eq:promotion_pge}
\end{equation}

The authors reparameterize these models in terms of $p_0$ to introduce covariate links for the cure fraction. Their generalized truncated Nadarajah--Haghighi specification for $S$ is classified under Axis~3.

\paragraph{Destructive exponentially weighted-Poisson construction.} 
The model of \cite{pal_roy_2026} has a different, treatment-mediated mechanism. In the simulation study, it first draws an initial number $M$ of risk factors from an exponentially weighted Poisson distribution,

\begin{equation}
P(M=m;\eta,\phi)=\exp\{-\eta e^{\phi}\}\frac{(\eta e^{\phi})^m}{m!},\qquad m=0,1,\ldots,
\label{eq:destructive_initial_count}
\end{equation}

and then treatment leaves each initial factor active with probability $p$. If $D$ denotes the number of active factors, then $D\mid M=m\sim\operatorname{Binomial}(m,p)$ and, with $W_j$ denoting the progression times,

\begin{equation}
Y=\begin{cases}\min\{W_1,\ldots,W_D\},&D>0,\\ \infty,&D=0.\end{cases}
\label{eq:destructive_lifetime}
\end{equation}

Under the exponentially weighted-Poisson specification, $D\sim\operatorname{Poisson}(\eta p e^{\phi})$, so the population survival and cure probability are

\begin{equation}
S_{\mathrm{pop}}(y)=\exp\{-\eta p e^{\phi}F(y)\},\qquad p_0=P(D=0)=\exp\{-\eta p e^{\phi}\}.
\label{eq:destructive_survival}
\end{equation}

Their progression-time specification is classified under Axis~3. Further, they use a sequential quadratic Hamiltonian estimation algorithm, which is discussed under Axis~5.

\paragraph{Integrated Poisson/negative-binomial promotion-time construction.}
Unlike cause-specific latent-count formulations, \cite{teh_et_al_2025} use one global latent count $M$ of unobserved competing causes. With $Z_1,\ldots,Z_M$ denoting the latent event times, $T=\min\{Z_1,\ldots,Z_M\}$ for $M>0$ and $T=\infty$ for $M=0$. If $A_M(s)=E(s^M)$ is the probability-generating function of $M$, then

\begin{equation}
S_p(t)=A_M\{S(t)\},\qquad p_0=P(M=0)=A_M(0),
\label{eq:teh_pgf}
\end{equation}

where $S(t)$ is the proper latent progression-time survival function. Rather than specifying a Bernoulli cure probability directly, they link the mean latent count $\theta(\bfx)=E(M\mid\bfx)$ to covariates through

\begin{equation}
\theta(\bfx)=\exp\{\eta(\bfx)\},\qquad \eta(\bfx)=\operatorname{act}\!\left(\operatorname{act}\!\left(\mathbf{B}_2^\top\operatorname{act}\!\left(\mathbf{B}_1^\top\bfx_i+\mathbf{b}_1\right)+\mathbf{b}_2\right)^\top\boldsymbol{\beta}_3\right),
\end{equation}

where $\eta(\cdot)$ is their convolutional-neural-network link function. Their integrated promotion-time cure model uses $M\sim\operatorname{Poisson}\{\theta(\bfx)\}$,

\begin{equation}
S_p(t\mid\bfx)=\exp{-\theta(\bfx)F(t)},\qquad p_0(\bfx)=\exp{-\theta(\bfx)},
\label{eq:teh_poisson}
\end{equation}

whereas their integrated negative-binomial cure model uses $M\sim\operatorname{NB}(1/\phi,1/[1+\phi\theta(\bfx)])$,

\begin{equation}
S_p(t\mid\bfx)=\{1+\phi\theta(\bfx)F(t)\}^{-1/\phi},\qquad p_0(\bfx)=\{1+\phi\theta(\bfx)\}^{-1/\phi}.
\label{eq:teh_nb}
\end{equation}

Thus, cure is the zero-count event $M=0$, and the negative-binomial extension accommodates dispersion through the count distribution rather than through separate cause-specific counts or dispersion variables.

\

The approach of \cite{kuttumannil_et_al_2020} is motivated by the promotion-time cure model of Tsodikov, but it does not introduce an explicit latent number $N$ of initiating causes. Instead, it models each cause-specific CIF through a semiparametric linear transformation model. It is therefore best described as promotion-time-related in motivation but CIF-based in its actual decomposition. We discuss its estimation under Axis~5. By contrast, \cite{ng_mclachlan_1998} uses an unobserved global cure mixture when competing-cause failures are sparse. However, its formulation does not specify a latent number of causes and is not itself a promotion-time construction. Its partial-likelihood treatment is considered under Axis~5.

\subsection{Defective (Improper) Survival Decomposition}
A fourth approach builds cure directly into an improper survival distribution, without introducing an explicit cure indicator, mixture weight, or latent count. In generic form,

\begin{equation}
S(t\mid \bfX)=p(\bfX)+\int_t^\infty f(u\mid \bfX)\,du,
\qquad
\lim_{t\to\infty}S(t\mid \bfX)=p(\bfX)>0,
\label{eq:defective}
\end{equation}

so that the total failure probability is

\begin{equation}
\int_0^\infty f(t\mid \bfX)\,dt=1-p(\bfX)<1.
\label{eq:defective_failure_probability}
\end{equation}

Thus, $p(\bfX)$ is induced by the parameter domain of the survival distribution itself, rather than being specified as a separate incidence parameter. In the competing-risks construction, assuming independence of the cause-specific survival factors, denoted by $S_j(t\mid \bfX)$, one gets

\begin{equation}
S(t\mid \bfX)=\prod_{j=1}^K S_j(t\mid \bfX),\qquad p_j(\bfX)=\lim_{t\to\infty}S_j(t\mid \bfX),\qquad p(\bfX)=\prod_{j=1}^K p_j(\bfX).
\label{eq:defective_competing}
\end{equation}

This construction is explicitly used by \cite{silpa_et_al_2024} and \cite{silpa_et_al_2024a}. The earlier paper, \cite{silpa_et_al_2024}, gives two distributional specifications. It links the scale and shape parameters for cause $j$ to covariates as $b_j(\bfx)=\exp(\bfx^\prime\bfbeta_j)$ and $a_j(\bfx)=\bfx^\prime\bfgamma_j$. Writing $a_j=a_j(\bfx)$ and $b_j=b_j(\bfx)$, in their defective Gompertz specification, $b_j>0$ and $a_j<0$, and

\begin{equation}
p_{\mathrm{G}}(\bfx)=\prod_{j=1}^K\exp\left\{\frac{b_j}{a_j}\right\}.
\label{eq:defective_gompertz}
\end{equation}

For their defective inverse-Gaussian specification, again with $b_j>0$ and $a_j<0$, and with $\Phi$ denoting the standard normal distribution function,

\begin{equation}
p_{\mathrm{IG}}(\bfx)=\prod_{j=1}^K\left(1-e^{2a_j/b_j}\right).
\label{eq:defective_inverse_gaussian}
\end{equation}

The two specifications differ in their latency distribution but have the same defective-survival decomposition, and the overall cure fraction is obtained by multiplying the cause-specific limiting survival probabilities. \cite{silpa_et_al_2024a} follows these same defective Gompertz and inverse-Gaussian constructions, with the change that the failure time is interval censored. The distributional details belong to Axis~3, whereas the treatment of interval censoring and estimation belong to Axes~4 and~5.

Unlike the zero-inflated mixture/CIF construction described above, defective-survival models obtain their cure fraction solely from the limiting tail of the specified survival distribution.


\section{Axis 3: Event-Time (Latency) Model Structure}
Axis~3 concerns the primary specification for the event-time component, which is the susceptible survival function $S_u(t\mid \bfX)$ in an incidence--latency model, or the corresponding latent-time, total-time, or cause-specific time component under another construction. Across the 26 reviewed papers, we observe the use of fully parametric or fixed-grid piecewise-constant models, PH models, AFT models, semiparametric transformation or CIF-based models, and nonparametric or only partially specified time structures. Below, a piecewise-constant baseline combined with a PH regression is classified as PH, and a fixed-grid piecewise-exponential progression-time model without a PH covariate component is classified as fully parametric. 

\subsection{Parametric and Piecewise-Constant Latency Models}
A fully parametric latency model fixes a finite-dimensional distributional family for a susceptible or latent failure time, for example,

\begin{equation}
S_{u,k}(t\mid \bfX)=S_{0,k}\{t;\vartheta_k(\bfX)\},\qquad t>0,
\label{eq:parametric_latency}
\end{equation}

where $S_{0,k}$ is drawn from a specified family. The earliest reviewed Weibull specification is that of \cite{greenhouse_wolfe_1984}, for which the susceptible disease-failure distribution has the usual form

\begin{equation}
F_{\mathrm{W}}(t;\alpha,\lambda)=1-\exp\{-\lambda t^\alpha\},\qquad \alpha>0,\ \lambda>0.
\label{eq:greenhouse_weibull_latency}
\end{equation}

\cite{chao_1998} instead use exponential relative-survival components, $S_r(t)=\exp(-\lambda_r t)$, for relapse and non-relapse-related death. \cite{choi_zhou_2002} retain the same finite-dimensional component-distribution strategy while considering exponential, Weibull, gamma, and piecewise-exponential families.

\cite{basu_tiwari_2010} formulate latency through cause- and cure-group-specific subhazards. Their fixed-grid specification is

\begin{equation}
h(r,t\mid q)=\sum_{d=1}^{D}\lambda_{qrd}I_{[t_{d-1},t_d)}(t),\qquad r=1,\ldots,R,
\label{eq:basu_piecewise_subhazard}
\end{equation}

and their parametric alternatives include exponential, Weibull, log-normal, log-logistic, gamma, and related log-time location--scale families. They compare equal- and ordered-hazard restrictions across cure groups, but the fitted alternatives remain finite-dimensional latency models.

The latent progression times in \cite{esmailian_et_al_2023} follow the generalized truncated Nadarajah--Haghighi family, which may be written $F_Z(t)=F_{\mathrm{GeTNH}}(t;\alpha,\beta,\lambda)$, and is given by
\[
F(t)
=
\frac{\exp\!\left\{1-(1+\lambda e^{-t^\beta})^\alpha\right\}
-\varrho(\alpha,\lambda)}
{1-\varrho(\alpha,\lambda)},
\]

where $\varrho(\alpha,\lambda)
=\exp\!\left\{1-(1+\lambda)^\alpha\right\}$.

\cite{silpa_et_al_2024} provide two cause-specific defective latency families. With $b_j(\bfx)=\exp(\bfx^\prime\bfbeta_j)$, $a_j(\bfx)=\bfx^\prime\bfgamma_j$, $b_j(\bfx)>0$, and $a_j(\bfx)<0$, their Gompertz survival is

\begin{equation}
S_{j,\mathrm{G}}(t\mid \bfx)=\exp\left\{-\frac{b_j(\bfx)}{a_j(\bfx)}\left[\exp\{a_j(\bfx)t\}-1\right]\right\}.
\label{eq:silpa_gompertz_latency}
\end{equation}

Their inverse-Gaussian survival is

\begin{equation}
S_{j,\mathrm{IG}}(t\mid \bfx)=1-\Phi\!\left(\frac{-1+a_j(\bfx)t}{\sqrt{b_j(\bfx)t}}\right)-\exp\!\left\{\frac{2a_j(\bfx)}{b_j(\bfx)}\right\}\Phi\!\left(\frac{-1-a_j(\bfx)t}{\sqrt{b_j(\bfx)t}}\right).
\label{eq:silpa_inverse_gaussian_latency}
\end{equation}

\cite{silpa_et_al_2024a} uses these same Gompertz and inverse-Gaussian latency families for its interval-censored extension.

\cite{ganguly_et_al_2026} use a distinct finite mixture of Weibull latency distributions. In their notation,

\begin{equation}
F_k(t)=1-\exp(-\lambda_k t^{\alpha_k}),\qquad S_{\mathrm{mix}}(t)=\sum_{k=1}^{K}\pi_kS_k(t),\qquad k=1,\ldots,K,
\label{eq:ganguly_weibull_mixture_latency}
\end{equation}

and, for a mode of interest, the covariate-linked Weibull parameter is $\lambda_{1l}(\bfx;\bfeta)=\exp(\bfeta^\top \bfx_l)$. \cite{pal_roy_2026} use the same two-parameter Weibull family in their fitted progression-time illustration. Their proposed semiparametric or nonparametric extension is not the fitted latency specification.

The neural-network component in \cite{teh_et_al_2025} parameterizes the latent-count stage rather than the latency distribution. Their progression time has a fixed-grid piecewise-exponential distribution: for $t\in(s_{g-1},s_g]$,

\begin{equation}
F(t;\boldsymbol\alpha)=1-\exp\left\{-\alpha_g(t-s_{g-1})-\sum_{j=1}^{g-1}\alpha_j(s_j-s_{j-1})\right\}.
\label{eq:teh_piecewise_exponential_latency}
\end{equation}

Here $S(t;\boldsymbol\alpha)=1-F(t;\boldsymbol\alpha)$. The model is therefore a piecewise-constant-hazard latency model, not a flexible or nonparametric latency model. Finally, \cite{sreedevi_et_al_2026} directly specify continuous cause-specific CIFs through a Gompertz family,

\begin{equation}
F_j^*(t_i\mid\theta_j)=1-\exp\left\{\frac{\lambda_j[1-\exp(\alpha_jt_i)]}{\alpha_j}\right\},\qquad \theta_j=(\alpha_j,\lambda_j).
\label{eq:sreedevi_gompertz_cif_latency}
\end{equation}

They link the parameters as $\alpha_j=\bfx_i^\prime\bfgamma_j<0$ and $\lambda_j=\exp(\bfx_i^\prime\bfvartheta_j)>0$.

Thus, its time component is fully parametric, although its target is a CIF rather than a susceptible survival function.

\subsection{Proportional-Hazards Latency Structure}
The earliest cause-specific PH formulation in the review is that of \cite{rejani_sankaran_2020},

\begin{equation}
\lambda_j(t\mid \bfz)=\lambda_{0j}(t)\exp(\bfbeta_j^\prime \bfz),\qquad j=1,\ldots,K,
\label{eq:ph_latency}
\end{equation}

where $\lambda_{0j}$ is the baseline cause-specific hazard. \cite{wang_et_al_2020a} and \cite{wang_et_al_2021} use the same PH latency structure for their two event-time components. \cite{menger_et_al_2023} likewise model the noncured cause-specific times through cause-specific hazards, using a piecewise-constant baseline representation, and \cite{wang_et_al_2026a} uses the same PH form with baseline survival functions estimated nonparametrically.

\cite{nicolaie_et_al_2019} use a distinct vertical PH specification. Conditional on susceptibility, their total event-time hazard and conditional relative-cause probabilities are

\begin{equation}
\lambda_\bullet(t\mid Y=1,\bfZ)=\lambda_0(t\mid Y=1)\exp(\bfgamma^\top \bfZ),
\label{eq:nicolaie_vertical_total_hazard}
\end{equation}

\begin{equation}
\pi_j(t\mid \bfU)=\frac{\exp\{\bfkappa_j^\top \bfB(t)+\bfupsilon_j^\top \bfU\}}{\sum_{l=1}^{J}\exp\{\bfkappa_l^\top \bfB(t)+\bfupsilon_l^\top \bfU\}},\qquad j=1,\ldots,J.
\label{eq:nicolaie_vertical_relative_cause}
\end{equation}

The PH component therefore governs the total susceptible event-time process, while a time-varying multinomial model assigns cause at failure. Thus, it is not a collection of separate ordinary Cox models for the individual causes.

\subsection{Accelerated Failure-Time Latency Structure}
The earliest reviewed AFT formulation is the cause-conditional semiparametric model of \cite{choi_et_al_2018}, which can be written as

\begin{equation}
\log T_{ij}=\bfX_i^\top\bfbeta_j+\varepsilon_{ij},\qquad j=1,\ldots,K,
\label{eq:aft_latency}
\end{equation}

with the cause-specific error distributions left unspecified. \cite{wang_et_al_2020b} use the same two-component semiparametric AFT structure, again leaving the error distributions unspecified. \cite{wang_et_al_2026b} retains the separate cause-specific AFT form but specifies the error distribution through Weibull, log-normal, or log-logistic alternatives, which makes it parametric within the AFT class.

\subsection{Semiparametric Transformation and CIF Models}
\cite{othus_et_al_2009} give the earliest general transformation model in the review,

\begin{equation}
S_g(t;\bar \bfZ_t)=g\left\{\int_0^t\exp\{\bfbeta^\top \bfZ(u)\}\,dH(u)\right\},
\label{eq:transformation_latency}
\end{equation}

where $H$ is a fixed but unspecified nondecreasing function with $H(0)=0$ and $g$ is a known continuous and strictly decreasing link function such that $g(0) = 1$ and $\lim_{t\to\infty} g(t) = 0$. The choices $g(x)=\exp(-x)$ and $g(x)=1/(1+x)$ yield time-dependent PH and proportional-odds submodels, respectively. \cite{choi_et_al_2015} apply this same class of semiparametric transformation models to the cause-conditional survival functions, with an unspecified monotone baseline function and a known link.

\cite{kuttumannil_et_al_2020} instead model each cause-specific CIF directly. With $F_k(t,J\mid \bfZ)=P(T\leq t,J=k\mid \bfZ)$, their linear transformation model is

\begin{equation}
g_k\{F_k(t,J\mid \bfZ)\}=h_k(t)+\bfZ^\top\bfbeta_k,\qquad k=1,\ldots,K,
\label{eq:cif_transformation_latency}
\end{equation}

where $g_k$ is a known increasing cause-specific link and $h_k$ is an unknown nondecreasing function. Proportional subdistribution-hazards and proportional-odds models are special cases. \cite{chen_et_al_2020} also take a CIF-based approach for acute vascular-access thrombosis, targeting $F_1(t\mid \bfX)=P(T\leq t,J=1\mid \bfX)$ with semiparametric regression that allows both time-varying and time-constant covariate effects. These papers model a CIF target directly rather than impose an ordinary PH model on a complete latent-time distribution.

\subsection{Nonparametric and Partially Specified Models}
\cite{li_et_al_2007} impose no parametric latency family or covariate regression; their population and susceptible survivals are estimated nonparametrically by the copula-graphic procedure described in Axis~5. \cite{ng_mclachlan_1998} formulate the cause-of-interest component only through $f_2(t;\theta_2)$ and $S_2(t;\theta_2)$ while using limited information from competing-cause failures. Because they do not specify a baseline family for this component, their latency structure is best described as partially specified rather than fully parametric or fully nonparametric.

\section{Axis 4: Dependence and Observation-Process Assumptions}
Axis~4 concerns dependence and observation-process assumptions that are not determined by the cure definition, joint-distribution decomposition, or latency model. It distinguishes dependence among latent cause-specific failure times or latent progression processes, the association of cure status with the event process, the relation between failure and censoring, including the censoring assumptions, within-cluster dependence, and the mechanism and treatment of missing or masked causes at an observed failure. These issues are logically separate---masked causes are a cause-observation problem rather than a form of stochastic dependence, and a model for failure--censoring dependence differs from a model for cross-cause dependence. In particular, a mixture over mutually exclusive event classes or a direct CIF model specifies the distribution of the observed event time and type but does not, by itself, identify a joint distribution for all counterfactual cause-specific failure times. Below, we classify the dependence and observation-process assumptions actually imposed by each paper. Latency specifications and estimation algorithms are considered under Axes~3 and~5, respectively.

\subsection{Joint Structure for Latent Risks and Cause Processes}
The earliest explicit conditional-independence construction is that of \cite{greenhouse_wolfe_1984}. Let $T_D^*$ and $T_O^*$ denote the latent times to disease death and other-cause death, respectively, and let $Q$ denote disease-cure status. Their construction has the conditional product form

\begin{equation}
P(T_D^*>t_D,T_O^*>t_O\mid Q,\bfX)=P(T_D^*>t_D\mid Q,\bfX)P(T_O^*>t_O\mid Q,\bfX).
\label{eq:conditional_independence_latent_risks}
\end{equation}

Cure changes the disease-death component, whereas the other-cause component is uncurable and does not depend on $Q$. Thus, this construction imposes independence between the two latent risks while linking cure status specifically to the disease process.

\cite{silpa_et_al_2024} assume the independent cause-specific defective components in Eq.~\eqref{eq:defective_competing}; \cite{silpa_et_al_2024a} retains the same cross-cause independence structure for interval-censored data. \cite{sreedevi_et_al_2026} likewise assumes independent causes, although its cause-specific CIF representation is a zero-inflated mixture/CIF model rather than a defective-survival construction.

\cite{esmailian_et_al_2023} consider conditional independence for unobserved failure-initiating causes. Given $M=m$ latent causes and heterogeneity variable $\eta$, their latent event times $R_1,\ldots,R_m$ satisfy

\begin{equation}
P(R_1>t_1,\ldots,R_m>t_m\mid M=m,\eta)=\prod_{j=1}^{m}S(t_j).
\label{eq:esmailian_conditional_independence}
\end{equation}

The latent count connects the number of active causes to cure and heterogeneity, but the model does not introduce a dependence parameter between progression times. \cite{teh_et_al_2025} follows the same conditional-i.i.d. principle for one global latent count $M$, and \cite{pal_roy_2026} does so for the progression times of its treatment-surviving active risk factors.

\cite{ganguly_et_al_2026} use the finite Weibull mixture in Eq.~\eqref{eq:ganguly_weibull_mixture_latency} to accommodate dependence and multimodality among failure modes. This is not a copula, shared-frailty, or other parameterization of a joint distribution for latent failure times $(T_1^*,\ldots,T_K^*)$.

\subsection{Models Without a Joint Cross-Cause Dependence Parameter}
Most of the remaining papers formulate cause membership and event time through mutually exclusive mixture classes, conditional cause-specific time distributions, direct CIFs, or a vertical factorization. These specifications model the observed pair $(T,D)$ but do not introduce an estimable association parameter for counterfactual latent times. The earliest such class-based formulation is \cite{choi_zhou_2002}, which assigns each subject to an immune or cause-specific category $B_i$ and then models the event time conditional on $B_i$. \cite{choi_et_al_2015}, \cite{choi_et_al_2018}, \cite{rejani_sankaran_2020}, \cite{wang_et_al_2020a}, \cite{wang_et_al_2020b}, \cite{wang_et_al_2021}, and \cite{wang_et_al_2026b} use the same basic event-class perspective, with different incidence and latency specifications. They should therefore not be described as assuming or estimating independence of latent competing failure times.

\cite{chao_1998} and \cite{ng_mclachlan_1998} likewise do not posit a joint distribution for all potential cause-specific failure times. 

The direct-CIF approaches also avoid a complete joint latent-time distribution. In \cite{kuttumannil_et_al_2020}, the overall survival is reconstructed from the cause-specific CIFs as

\begin{equation}
S(t\mid \bfZ)=1-\sum_{k=1}^{K}F_k(t\mid \bfZ).
\label{eq:kuttumannil_cif_survival}
\end{equation}

\cite{chen_et_al_2020} similarly models the cure indicator and only the selected cause-of-interest CIF, rather than a joint distribution of cure status, failure time, and all failure types. The vertical factorization of \cite{nicolaie_et_al_2019} is also distinct from a latent-time dependence model as it assigns cause conditionally at the event time and avoids a baseline joint distribution for future cause labels. Finally, \cite{wang_et_al_2026a} retains the marginal mixture/event-class construction of the Wang noncurable-risk models, whereas its additional within-cluster working correlations are a separate dependence feature described below.

\subsection{Association of Cure Status With the Event Process}
Cure can be linked to an event process without specifying stochastic dependence among latent risks. The Greenhouse--Wolfe construction above links disease cure to the disease-death component while leaving the other-cause process unaffected. \cite{basu_tiwari_2010} give the earliest explicit cause-specific cure-group hazard formulation. With $Q=1$ for the cured group and primary cause $C=1$, they impose

\begin{equation}
h(C=1,t\mid Q=1)=0,\qquad t\geq 0.
\label{eq:basu_primary_cure_hazard}
\end{equation}

For nonprimary causes, their general model allows $h(C=r,t\mid Q=1)$ and $h(C=r,t\mid Q=0)$ to differ. Their equal-hazards restriction is

\begin{equation}
h(C=r,t\mid Q=1)=h(C=r,t\mid Q=0)=h(r,t),\qquad r=2,\ldots,R,
\label{eq:basu_equal_hazards}
\end{equation}

whereas their ordered-hazards restriction is

\begin{equation}
\sum_{r=2}^{R}h(r,t\mid Q=1,\theta)\leq\sum_{r=1}^{R}h(r,t\mid Q=0,\theta),\qquad t\geq 0.
\label{eq:basu_ordered_hazards}
\end{equation}

Thus, Basu and Tiwari permit cure status to alter the nonprimary-cause hazards or impose structured equality/order constraints, and hence, this is a cure--event-process association.

\cite{menger_et_al_2023} has a different cause-specific cure construction, with a latent cure indicator for each cause and a noncured survival process for that cause. It therefore associates cause-specific cure status with its corresponding event process, but does not introduce a shared frailty or a cross-cause association parameter. In \cite{esmailian_et_al_2023}, \cite{teh_et_al_2025}, and \cite{pal_roy_2026}, cure is instead tied structurally to the zero count of latent or active risks. The count mechanisms are classified under Axis~2. These mechanisms create latent heterogeneity in the event process but not an estimable pairwise dependence parameter among observed competing causes.

\subsection{Within-Cluster Dependence}
The clustered marginal framework of \cite{wang_et_al_2026a} is the only reviewed model that explicitly introduces working correlation parameters for dependence between individuals. For cluster $i$, their four working correlation matrices are

\begin{equation}
\bfQ_i(\rho_h)=\{q_{jj'}(\rho_h)\}_{n_i\times n_i},\qquad h=1,\ldots,4,
\label{eq:wang_working_correlation}
\end{equation}

and their exchangeable specification is

\begin{equation}
q_{jj'}(\rho_h)=1\ (j=j'),\qquad q_{jj'}(\rho_h)=\rho_h\ (j\ne j'),\qquad h=1,\ldots,4.
\label{eq:wang_exchangeable_correlation}
\end{equation}

The four matrices represent working associations for susceptibility, event class, and the two event-time components. Exchangeable and AR(1) structures are considered. These are working-correlation parameters for within-cluster cure statuses, event statuses, and failure-time contributions.

\subsection{Failure--Censoring Dependence}
The earliest reviewed explicit joint model for a failure time and censoring time is \cite{li_et_al_2007}. For the possibly improper failure time $T$ and censoring time $C$, they use an Archimedean copula,

\begin{equation}
P(T>t,C>u)=C_\theta\{S_T(t),S_C(u)\}.
\label{eq:li_failure_censoring_copula}
\end{equation}

This is a dependence model for failure and censoring, not for competing failure causes. The specified copula supplies the otherwise unidentifiable association structure.

\cite{othus_et_al_2009} and \cite{chen_et_al_2020} handle dependent censoring through IPCW estimating equations, described in Axis~5, without imposing a parametric copula or estimating a full joint distribution of censoring and competing failure times.

Other reviewed models use noninformative or conditionally independent censoring rather than modeling a failure--censoring association. For example, \cite{wang_et_al_2020b} write

\begin{equation}
C_i\perp(T_{1,i},T_{2,i},Y_i)\mid \bfX_i.
\label{eq:wang_independent_censoring}
\end{equation}

\cite{wang_et_al_2026a} uses the same condition with cluster indices. The likelihood formulations in \cite{choi_et_al_2015, choi_et_al_2018, esmailian_et_al_2023, sreedevi_et_al_2026} likewise use the usual noninformative or conditionally independent-censoring convention, while \cite{ganguly_et_al_2026} and \cite{pal_roy_2026} use random noninformative right censoring. Interval censoring in \cite{silpa_et_al_2024a} and \cite{wang_et_al_2026b} changes the observation scheme, but is not itself a failure--censoring dependence model.

\subsection{Masked Causes and Latent Cause Indicators}
Masked causes are missing cause labels at an observed failure, rather than stochastic dependence between risks. The earliest reviewed treatment is \cite{basu_tiwari_2010}. If $C=R+1$ denotes a recorded death with an unknown true cause among $1,\ldots,R$, they marginalize its contribution as

\begin{equation}
p(C=R+1,t)=\sum_{r=1}^{R}p(C=r,t).
\label{eq:basu_masked_cause}
\end{equation}

Writing $\mathcal M=1$ when the cause is masked, their formulation further makes the masking-symmetry assumption

\begin{equation}
P(\mathcal M=1\mid C=r_1,T=t)=P(\mathcal M=1\mid C=r_2,T=s),
\label{eq:basu_masking_symmetry}
\end{equation}

for all $r_1,r_2\in\{1,\ldots,R\}$ and $s,t>0$. Thus, the probability of masking does not depend on the true cause or failure time. \cite{menger_et_al_2023} likewise integrates masked-cause observations over their feasible cause set and augments the model with cause-specific cure indicators, but it does not impose the same simple binary cure-group structure as Basu and Tiwari. In \cite{chao_1998}, by contrast, the latent variables for censored subjects are cure and relapse indicators, not masked causes at an observed failure.

\section{Axis 5: Estimation Framework}
Axis~5 concerns how each model is fitted and how its uncertainty or predictive performance is assessed. Let $\mathcal{O}_i=(\widetilde T_i,\delta_i,D_i,\bfX_i)$ denote the observed data for subject $i$, where $\delta_i=1$ indicates a failure and $D_i$ is its observed cause. For right-censored data, a common observed-data log-likelihood template is

\begin{equation}
\ell_n(\bfvartheta)=\sum_{i=1}^{n}\{(1-\delta_i)\log S_i(\widetilde T_i;\bfvartheta)+\delta_i\log f_{D_i,i}(\widetilde T_i;\bfvartheta)\},\qquad \widehat\bfvartheta=\arg\max_{\bfvartheta}\ell_n(\bfvartheta),
\label{eq:axis5_observed_likelihood}
\end{equation}

where, $S_i$ and $f_{D_i,i}$ abbreviate the model-specific population survival and observed-cause density conditional on covariates $\bfX_i$, and $\bfvartheta$ may contain finite-dimensional parameters and, in semiparametric models, unknown functions. Equation~\eqref{eq:axis5_observed_likelihood} is only a fitting template---interval-censored, masked-cause, defective, copula, and latent-count models replace the relevant likelihood contributions with their own observed-data probabilities. Their structural ingredients are classified under Axes~1--4.

\subsection{Direct and Partial Likelihood Maximization}
The earliest reviewed direct-likelihood implementation is \cite{greenhouse_wolfe_1984}, which maximizes the likelihood for the cure probability and Weibull parameters with an iterative quasi-Newton procedure. \cite{choi_zhou_2002} follows the same direct observed-likelihood principle for its parametric mixture model, establishes existence, consistency, and asymptotic normality of the maximum-likelihood estimators, and derives likelihood-ratio tests. By contrast, \cite{ng_mclachlan_1998} maximizes a partial likelihood for the cause-of-interest component,

\begin{equation}
\widehat\bfvartheta_1=\arg\max_{\bfvartheta_1}\ell_{\mathrm{partial}}(\bfvartheta_1),
\label{eq:ng_partial_likelihood}
\end{equation}

so that sparse competing-cause failures need not be represented by a fully specified joint likelihood. It is likelihood-based, but it is not a full direct likelihood for all causes.

Later fully parametric models retain the maximization principle in \eqref{eq:axis5_observed_likelihood}, while substituting their own population survival and density functions. \cite{esmailian_et_al_2023} use direct maximum likelihood under right censoring for the latent-count, GeTNH, and regression parameters, with AIC and BIC for comparison. \cite{silpa_et_al_2024} fit the defective inverse-Gaussian and Gompertz models by BFGS optimization and use the observed information for standard errors and confidence intervals. \cite{silpa_et_al_2024a} follows the same likelihood principle for interval-censored contributions, and reports inverse-information uncertainty estimates and AIC/BIC. \cite{sreedevi_et_al_2026} uses constrained numerical maximum likelihood for its zero-inflation, cure, and Gompertz-regression parameters and its Hessian for standard errors and asymptotic confidence intervals. \cite{pal_roy_2026} also maximizes the observed likelihood directly, but uses the gradient-free sequential quadratic Hamiltonian (SQH) algorithm rather than a conventional optimizer, comparing SQH with conjugate-gradient line search, EM, and standard R optimizers. Thus, its distinctive computational contribution remains a specialized direct-likelihood optimizer.

\subsection{Semiparametric and Nonparametric Likelihood}
The earliest semiparametric likelihood procedure in the review is \cite{choi_et_al_2015}. With $\mathcal H$ denoting the unspecified transformation component, their nonparametric maximum-likelihood estimator has the form

\begin{equation}
(\widehat\bfvartheta,\widehat{\mathcal H})=\arg\max_{\bfvartheta,\mathcal H\in\mathcal A}\ell_n(\bfvartheta,\mathcal H),
\label{eq:choi2015_npml}
\end{equation}

where $\mathcal A$ is the admissible function space. Their joint parametric and nonparametric likelihood combines multinomial incidence regression with cause-specific transformation models---martingale integral representations yield the score equations, and the inverse observed information supplies variance estimates for the regression parameters. \cite{choi_et_al_2018} follow the same semiparametric likelihood/profile-likelihood principle, but replace the transformation component with cause-specific AFT models having unspecified error distributions. They estimate these distributions through kernel smoothing and base inference on a profile likelihood smoothed by a kernel. Neither procedure is classified as a separate EM algorithm.

\subsection{EM Algorithms for Latent Cure and Event States}
For a latent cure, susceptibility, or event-state variable $Z_i$, an EM iteration takes the generic form

\begin{equation}
Q(\bfvartheta\mid\bfvartheta^{(m)})=E_{\bfvartheta^{(m)}}\{\ell_c(\bfvartheta;\mathcal{O},\bfZ)\mid \mathcal{O}\},\qquad \bfvartheta^{(m+1)}=\arg\max_{\bfvartheta}Q(\bfvartheta\mid\bfvartheta^{(m)}),
\label{eq:generic_em}
\end{equation}

where $\ell_c$ is the complete-data log-likelihood and $\bfvartheta^{(m)}$ is the estimated parameter vector at iteration $m$. In \cite{nicolaie_et_al_2019}, the latent susceptibility status is handled by EM for the cure and total-failure components, while the relative-cause component is fitted with multinomial regression and time splines. \cite{rejani_sankaran_2020} follows the same latent-state likelihood strategy, updating posterior susceptibility weights, incidence coefficients, Cox regression coefficients, and baseline cumulative hazards in its E- and M-steps.

\cite{wang_et_al_2020b} uses EM because cure and event indicators are partly latent. Their M-step maximizes a kernel-smoothed conditional profile likelihood with a piecewise-constant hazard approximation. \cite{wang_et_al_2021} likewise use EM-based semiparametric maximum likelihood and develop asymptotic theory for the parametric estimators. \cite{ganguly_et_al_2026} use EM under right censoring, with latent immune/susceptible status and mixture-mode membership in the E-step and cure, mixing, Weibull, and covariate-link parameters in the M-step. They report confidence intervals and conditional survival predictions for patients alive at a given time. \cite{wang_et_al_2026b} use the same EM principle for interval- and right-censored observations, computing posterior cure/event-category probabilities and then updating multinomial incidence and cause-specific AFT parameters for Weibull, log-normal, or log-logistic latency distributions.

\subsection{EM With a Neural-Network M-Step and Profile Likelihood}
The neural latent-count model of \cite{teh_et_al_2025} is also an EM-based likelihood procedure, rather than a standalone direct neural-network fit. With complete data $\bfD_{\mathrm{comp}}=(\bft,\bfdelta,\bfx,\bfM)$, neural-network weights $\bfw$, and latent-time parameters $\bfalpha$, their complete-data likelihood is

\begin{equation}
L(\bfpsi;\bfD_{\mathrm{comp}}) = \prod_{i=1}^{n}[m_i f(t_i;\bfalpha)]^{\delta_i}S(t_i;\bfalpha)^{m_i-\delta_i}P_{\Theta_i}(M_i=m_i),\qquad \bfpsi=(\bfw,\bfalpha).
\label{eq:teh_complete_likelihood}
\end{equation}

Their E-step calculates

\begin{equation}
m_i^{(k+1)}=E(M_i+\delta_i\mid \bfD_{\mathrm{obs}},\bfalpha^{(k)},\bfw^{(k)}),
\label{eq:teh_estep}
\end{equation}

and the M-step separately maximizes the component involving $\bfw$ through the convolutional neural network and the component involving $\bfalpha$ for the latent-time distribution. For the negative-binomial version, the overdispersion parameter is selected by profile log-likelihood. The neural optimization is therefore contained within the M-step. The fitted models are evaluated through loss trajectories, AUC, and pointwise confidence intervals for estimated survival curves.

\subsection{Estimating Equations and IPCW-Based Inference}
The earliest reviewed approach that avoids a fully specified likelihood is \cite{othus_et_al_2009}. Their inverse-censoring-probability weighted estimating-equation principle can be represented schematically as

\begin{equation}
U_n^{\mathrm{IPCW}}(\bfvartheta)=\sum_{i=1}^{n}\frac{\psi_i(\bfvartheta)}{\widehat G_C^{\mathrm{crude}}(\widetilde T_i-\mid \bfX_i)}=0,
\label{eq:othus_ipcw}
\end{equation}

where $\widehat G_C^{\mathrm{crude}}$ is estimated from the crude censoring process. They derive unbiased equations for the cure, latency, and transformation-function parameters with time-dependent covariates and dependent censoring, solve them iteratively, and use a weighted bootstrap for asymptotic variance estimation.

\cite{chen_et_al_2020} follow the same IPCW principle but uses a two-stage procedure, schematically

\begin{equation}
U_{\mathrm{cure}}(\bfgamma)=0,\qquad U_{\mathrm{CIF}}(\bfbeta;\widehat\bfgamma)=0,
\label{eq:chen_two_stage_ipcw}
\end{equation}

for logistic cure-regression coefficients $\bfgamma$ and selected-CIF parameters $\bfbeta$. Thus, theirs is an IPCW/estimating-equation method rather than an EM algorithm. \cite{kuttumannil_et_al_2020} likewise use counting-process estimating equations for regression coefficients and unknown transformation functions, with martingale large-sample theory instead of a complete likelihood.

The clustered method of \cite{wang_et_al_2026a} uses an expectation-solution (ES) algorithm, where after an E-step for unobserved susceptibility and event states, an S-step solves correlation-adjusted estimating equations of the schematic GEE form

\begin{equation}
\sum_{i=1}^{n}\mathbf D_i^{\mathsf T}\mathbf V_i^{-1}(\mathbf Y_i-\boldsymbol\mu_i)=\mathbf 0,
\label{eq:wang2026a_gee}
\end{equation}

where $\mathbf D_i$ is a derivative matrix and $\mathbf V_i$ incorporates the working correlation. The S-step uses Newton--Raphson updates, estimates nonparametric baseline survival functions, and updates the working correlations by moment equations. Cluster-level bootstrap resampling provides variance estimates. This is an EM/estimating-equation hybrid for clustered data, not ordinary independent-sample EM.

\subsection{Bayesian Posterior Computation and Data Augmentation}
The earliest Bayesian fitting framework in the review is \cite{chao_1998}. With latent variables $Z$ included in the augmented data, the posterior has the standard form

\begin{equation}
p(\bfvartheta,\bfZ\mid \mathcal{O})\propto L_c(\bfvartheta;\mathcal{O},\bfZ)p(\bfvartheta).
\label{eq:bayesian_augmented_posterior}
\end{equation}

Chao samples this posterior via Gibbs sampling, imputing cure and relapse indicators for censored subjects, where cure and event-type probabilities receive beta priors, and exponential-rate parameters receive gamma priors. \cite{basu_tiwari_2010} extends Bayesian augmentation to masked causes and latent cure status, using a block sampler with random-walk Metropolis updates for hazard parameters and full-conditional updates for the cure fraction and latent indicators. Bayes factors compare their general, equal-hazard, and ordered-hazard specifications.

\cite{wang_et_al_2020a} follows the same posterior-augmentation principle for latent cure and event indicators, assigns multivariate-normal priors to regression parameters and a gamma-process prior to cumulative baseline hazards, and uses MCMC with Gibbs and adaptive-rejection updates. \cite{menger_et_al_2023} likewise sample an augmented posterior for cause-specific cure indicators, noncured failure times, and masked causes. They use gamma priors for piecewise-constant baseline hazards, specified priors for cure and hazard coefficients, and a combination of Gibbs, adaptive-rejection, and localized Metropolis updates. Cause-specific DIC and C-index measures are used for model assessment. Although these methods augment latent data as EM does, they sample a posterior distribution rather than maximize the $Q$ function in \eqref{eq:generic_em}.

\subsection{Copula-Graphic Nonparametric Estimation}
\cite{li_et_al_2007} use a copula-graphic procedure. After constructing a copula-graphic estimate $\widehat S_{\mathrm{CG}}$ of the population survival under a specified Archimedean failure--censoring copula, they obtain the cure fraction and uncured survival as

\begin{equation}
\widehat\pi_{\mathrm{cure}}=\lim_{t\to\infty}\widehat S_{\mathrm{CG}}(t),\qquad \widehat S_u(t)=\frac{\widehat S_{\mathrm{CG}}(t)-\widehat\pi_{\mathrm{cure}}}{1-\widehat\pi_{\mathrm{cure}}}.
\label{eq:li_copula_graphic}
\end{equation}

They establish consistency, asymptotic normality, weak convergence, and hypothesis tests for cure fractions and latency distributions. This work is neither a finite-dimensional maximum likelihood nor an IPCW estimating-equation method. Their identifying assumption is the specified copula, classified under Axis~4.

\section{Software and Reproducibility}
Software availability is limited and uneven across the 26 reviewed contributions (Table~\ref{tab:software_summary}). No reviewed paper provides a clearly licensed, widely adopted R or Python package implementing its competing-risk cure method. Public access to an article, repository, or code appendix should not be equated with open-source software, which requires an explicit license.

\begin{table}[htbp]
\centering
\caption{Reported software availability in the 26 reviewed competing-risk cure papers}
\label{tab:software_summary}
\footnotesize
\setlength{\tabcolsep}{2.5pt}
\renewcommand{\arraystretch}{0.95}
\begin{tabular}{@{}p{0.27\textwidth}p{0.63\textwidth}@{}}
\hline
\textbf{Availability ($n$)} & \textbf{Reported evidence and practical interpretation} \\
\hline
Public author repository (3) & Complete GitHub code is reported by \cite{teh_et_al_2025}, \cite{menger_et_al_2023}, and \cite{pal_roy_2026}; the reported environments are Python/TensorFlow, FORTRAN~95/IMSL, and R, respectively. No article reports an explicit software license. \\
Analysis code in article appendix (1) & \cite{silpa_et_al_2024} include melanoma-data analysis code in the article appendix, but report neither a dedicated repository nor a software license. \\
Named tools, packages, or comparator software only (9) & The papers identify packages, numerical routines, general software, or comparator programs, but no complete author-provided implementation or software license. \\
No named or verifiable source release (13) & The papers report neither an author repository nor a code archive sufficient for independent verification; broad references to software or an unnamed program do not constitute a reusable implementation. \\
\hline
\end{tabular}
\par\smallskip
{\footnotesize\textit{Note.} Categories are mutually exclusive and summarize information reported in the papers. The \textit{named tools} category includes general packages, numerical routines, and comparator software; it does not indicate that the proposed method is available as reusable code.}
\end{table}

The three repository-level reports are the strongest code-availability evidence. \cite{teh_et_al_2025} implemented their model in Python in Google Colab Pro, using TensorFlow~2.9.2 for the convolutional neural network, Keras data-loading utilities, and \texttt{lifelines}, including \texttt{KaplanMeierFitter}; their complete implementation is linked at \url{https://github.com/TehLedRed/SpringerManuscript}. \cite{menger_et_al_2023} report FORTRAN~95 with IMSL subroutines and link complete code at \url{https://github.com/austinmenger/Bayesian_CompRisk_Cure_Masked}; \cite{pal_roy_2026} report R simulations and link code at \url{https://github.com/suvrapal/SQH-EWP}. The reviewed articles do not report an explicit software license for any of these repositories, so they are best described as publicly accessible code rather than definitively open-source software. \cite{silpa_et_al_2024} paper is the only other paper with author-provided code: its appendix contains melanoma-data analysis code using \texttt{optim} from \texttt{stats}, the melanoma data set from \texttt{timereg}, and \texttt{smcure} for a proportional-hazards mixture-cure comparison, but it provides no dedicated repository or software license.

The nine named-tool reports document computational ingredients rather than access to complete author implementations. \cite{silpa_et_al_2024a} use \texttt{nlminb}; their illustration uses \texttt{pseudo.hiv.long} from \texttt{intccr} and \texttt{ic\_np} from \texttt{icenReg} for a Turnbull-estimator plot. \cite{sreedevi_et_al_2026} use \texttt{optim} from \texttt{stats}, \texttt{flexsurvreg} from \texttt{flexsurv}, and \texttt{survfit} from \texttt{survival}; \cite{nicolaie_et_al_2019} mention SAS \texttt{PROC GLM}, \texttt{glm}, and \texttt{smcure}. \cite{esmailian_et_al_2023} cite the \texttt{timereg} and \texttt{survival} package webpages and mention Mathematica \texttt{NMaximize} and \texttt{bbmle}; \cite{wang_et_al_2021} use \texttt{smcure} only for comparison; and \cite{basu_tiwari_2010} use CANSURV as a comparator, noting that it does not handle competing risks. \cite{choi_zhou_2002} use MATLAB, \cite{ng_mclachlan_1998} use a FORTRAN program with NAG routines \texttt{GO5CAF} and \texttt{GO5CBF} for pseudo-random-number generation, and \cite{wang_et_al_2026b} mention \texttt{numDeriv} as an optional resource. None releases a complete, dedicated implementation or reports a software license for the proposed method.

For the 13 papers in the final row of Table~\ref{tab:software_summary}, no reusable implementation can be verified from the reported information. \cite{wang_et_al_2020a} refer to an unnamed statistical programming environment and existing popular packages for semiparametric Bayesian analysis, whereas \cite{wang_et_al_2020b} refer only to unnamed popular packages; \cite{choi_et_al_2015} similarly mention scientific-computing packages only generically. \cite{kuttumannil_et_al_2020} mention a program used for computation but provide neither source code nor public access, and \cite{rejani_sankaran_2020} and \cite{chen_et_al_2020} name no package, environment, or repository. The other no-release papers---\cite{greenhouse_wolfe_1984, chao_1998, li_et_al_2007, othus_et_al_2009, choi_et_al_2018}---provide no named software or code-access statement. 
The absence of standardized, clearly licensed, and verifiable implementations remains a barrier to independent replication, systematic benchmarking, and routine methodological adoption.

\section{Illustrative Real-Data Application}
\label{sec:bmt_application}

For this illustration, we applied the defective Gompertz competing-risks cure construction of \cite{silpa_et_al_2024} to the public \texttt{bmt} data set distributed with the \texttt{geecure} R package. The data contain 137 patients with acute leukemia who underwent bone-marrow transplantation. Disease-free survival time was defined from transplantation to the first of relapse, death, or the end of follow-up. We constructed a first-event competing-risk outcome in which relapse was cause 1, death before relapse was cause 2, and patients alive and relapse-free at the end of follow-up were right-censored. Thus, cause 2 is not interpreted as treatment-related or other-cause mortality. Time was expressed in years. We compared patients with high-risk acute myeloid leukemia (AML), coded $x=1$, with patients with acute lymphoblastic leukemia or low-risk AML, coded $x=0$. Table~\ref{tab:bmt_data_summary} summarizes the sample size, event frequencies, observation time, and disease-risk-group distribution.

\begin{table}[htbp]
\centering
\caption{Summary of the bone-marrow-transplant data used in the illustrative analysis. Median observation time is measured from transplantation to the first event or censoring.}
\label{tab:bmt_data_summary}
\begin{tabular}{lrlrlr}
\hline
$N$ & 137 & Relapses & 42 & Deaths before relapse & 41 \\
\hline
Censored & 54 & Median observation time (years) & 1.28 & AML high risk & 45 (32.8\%) \\
\hline
\end{tabular}
\end{table}

\subsection{Model and Fitting}
For cause $j\in\{1,2\}$, we specified the defective Gompertz hazard $\lambda_j(t\mid x)=b_j(x)\exp\{a_j(x)t\}$, where $b_j(x)>0$ and $a_j(x)<0$. The corresponding cause-specific survival and overall event-free survival are

\begin{equation*}
S_j(t\mid x)=\exp\left[\frac{b_j(x)}{a_j(x)}\{1-\exp(a_j(x)t)\}\right],\qquad S(t\mid x)=S_1(t\mid x)S_2(t\mid x).
\label{eq:bmt_defective_gompertz}
\end{equation*}

The component-specific limiting survival probabilities and the global event-free tail probability are

\begin{equation*}
p_j(x)=\lim_{t\to\infty}S_j(t\mid x)=\exp\left\{\frac{b_j(x)}{a_j(x)}\right\},\qquad p(x)=p_1(x)p_2(x).
\label{eq:bmt_limiting_survival}
\end{equation*}

Within this first-event formulation, $p_1(x)$ is the limiting probability of avoiding the relapse component, $p_2(x)$ is the corresponding probability for the death-before-relapse component, and $p(x)$ is the model-derived global event-free cure probability. The product representation assumes independent cause-specific defective components. To enforce the defective domain for both risk groups, we used $a_j(x)=-\exp(\gamma_{j0}+\gamma_{j1}x)$ and $b_j(x)=\exp(\beta_{j0}+\beta_{j1}x)$. The observed-data log-likelihood was maximized under independent right censoring using multi-start L-BFGS-B optimization. For a failure of type $j$, the contribution is $\log\{S(t\mid x)\lambda_j(t\mid x)\}$, whereas a censored observation contributes $\log S(t\mid x)$. Model-based CIFs were evaluated as $F_j(t\mid x)=\int_0^t S(u\mid x)\lambda_j(u\mid x)\,du$ by numerical integration.

We fitted an intercept-only model (M0) and a model allowing high-risk AML status to affect both $a_j(x)$ and $b_j(x)$ for each cause (M1). We used 200 nonparametric bootstrap resamples to obtain percentile intervals for the limiting survival probabilities. The empirical event-free survival and cumulative-incidence curves are Kaplan--Meier (Figure~\ref{fig:bmt_event_free_survival}) and Aalen--Johansen estimates (Figure~\ref{fig:bmt_cif}), respectively, and are included as descriptive benchmarks rather than nonparametric estimators of cure.

\begin{table}[htbp]
\centering
\caption{Likelihood comparison of defective Gompertz models for the bone-marrow-transplant data. M1 allows high-risk AML status to affect both Gompertz parameters in each cause component.}
\label{tab:bmt_model_fit}
\begin{tabular}{lrrrr}
\hline
Model & Parameters & Log-likelihood & AIC & BIC \\
\hline
M0: no disease-risk-group effect & 4 & -209.91 & 427.82 & 439.50 \\
M1: AML high-risk effect & 8 & -202.52 & 421.04 & 444.40 \\
\hline
\end{tabular}
\end{table}

\subsection{Results}
Table~\ref{tab:bmt_model_fit} shows the likelihood, AIC, and BIC for the intercept-only and high-risk AML models. Relative to M0, M1 increased the log-likelihood by 7.39 and reduced the AIC by 6.78, but its BIC was 4.90 higher because it estimated four additional parameters. Thus, the AIC supports modeling the risk-group contrast, whereas the BIC favors the more parsimonious model. We nevertheless present the M1 estimates to illustrate the model's group-specific quantities rather than to claim a definitive model-selection result.

\begin{figure}[t]
\centering
\includegraphics[width=0.88\textwidth]{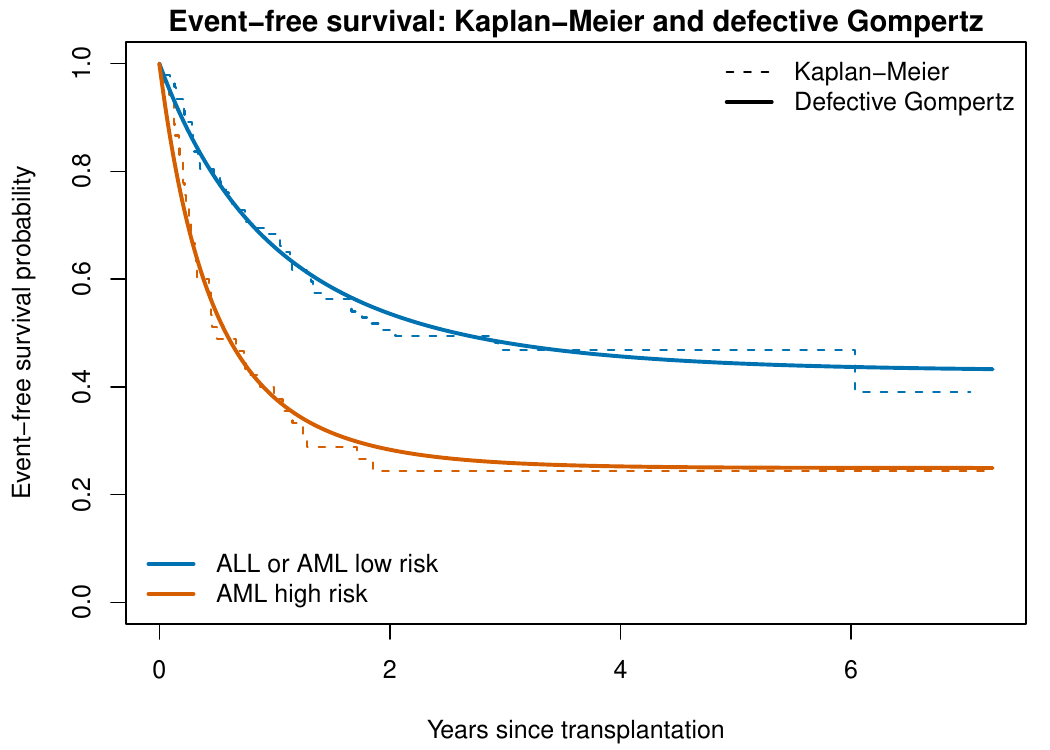}
\caption{Empirical event-free survival curves (thin dashed Kaplan--Meier curves) and fitted defective Gompertz survival curves (thick solid curves), stratified by disease-risk group.}
\label{fig:bmt_event_free_survival}
\end{figure}

\begin{figure}[htbp]
\centering
\includegraphics[width=0.88\textwidth]{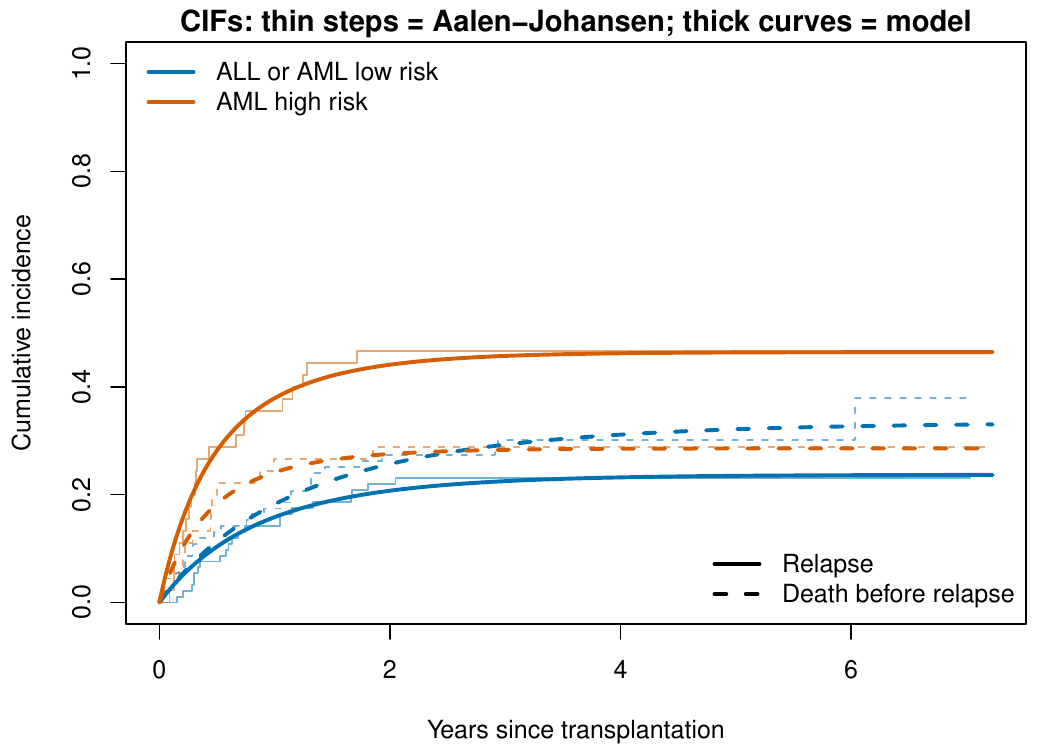}
\caption{Empirical cause-specific cumulative-incidence functions (thin Aalen--Johansen step functions) and fitted defective Gompertz cumulative-incidence functions (thick curves). Colors denote disease-risk group, solid curves denote relapse, and dashed curves denote death before relapse.}
\label{fig:bmt_cif}
\end{figure}

Figure~\ref{fig:bmt_event_free_survival} shows more rapid loss of event-free survival among patients with high-risk AML. The fitted curves follow the broad pattern of the Kaplan--Meier estimates and approach group-specific positive tails. Figure~\ref{fig:bmt_cif} shows that the group contrast is most pronounced for relapse, with a higher fitted and empirical CIF among high-risk AML patients. The fitted death-before-relapse curves show comparatively less separation.

Table~\ref{tab:bmt_cure} reports the model-derived tail probabilities. The relapse-component tail probability was lower for high-risk AML than for the comparison group, with estimates of 0.418 and 0.715, respectively. The death-before-relapse component estimates were nearly identical at 0.596 and 0.600. Consequently, the global event-free tail estimate was lower for high-risk AML at 0.249 than for the comparison group at 0.429. The bootstrap intervals quantify substantial uncertainty and overlap for the global estimates.

\begin{table}[htbp]
\centering
\caption{Model-derived component-specific and global limiting survival probabilities. Entries are estimates (percentile bootstrap 95\% intervals; $B=200$). $p_1$ denotes the limiting survival probability for the relapse component, $p_2$ denotes the corresponding probability for the death-before-relapse component, and $p=p_1p_2$ is the global event-free tail probability.}
\label{tab:bmt_cure}
\begin{tabular}{lccc}
\hline
Disease group & $p_1$: relapse & $p_2$: death before relapse & $p$: global event-free \\
\hline
ALL or AML low risk & 0.715 (0.593, 0.831) & 0.600 (0.359, 0.732) & 0.429 (0.232, 0.536) \\
AML high risk & 0.418 (0.242, 0.604) & 0.596 (0.432, 0.764) & 0.249 (0.136, 0.381) \\
\hline
\end{tabular}
\end{table}

The group contrast in the global event-free tail arises primarily from the relapse component rather than from the death-before-relapse component. These quantities remain model-defined extrapolations. The data include only 137 patients, the median observed time is 1.28 years, and the positive tails depend on the restrictive defective Gompertz form, the assumed independence of the two cause components, and independent censoring. Accordingly, the application demonstrates how a defective-survival competing-risk cure model can be fitted and interpreted in R, but the tail probabilities should not be treated as definitive clinical cure estimates.

\section{Discussion and Conclusion}
This review shows that competing-risk cure models are not a single model class: apparently similar models can differ in cure estimand, joint-distribution construction, event-time target, observation assumptions, regression interpretation, and identifying conditions. The five-axis framework makes these distinctions explicit and separates structural choices from fitting choices.

Across the review, the central finding is that a positive limiting survival probability does not identify a common cure estimand. It may represent global event-free cure, cure from one event while competing events remain possible, separate cause-specific cure, an explicit mixture mass, a zero count of active latent causes, or the improper tail of a defective distribution. These quantities have different scientific meanings and require explicit reporting of the noncurable competing risks, cure definition, and tail assumptions. A flexible covariate link should also not be mistaken for a flexible latency structure.

Likewise, a model for observed event classes or cause-specific CIFs does not by itself define a joint distribution for counterfactual latent failure times. Explicit cross-cause dependence remains uncommon, and failure--censoring dependence, within-cluster association, and masked causes are distinct observation-process issues. Estimation methods are not structural labels: direct likelihood, EM, Bayesian augmentation, IPCW estimating equations, and copula-graphic estimation control how latent states and nuisance components are handled, but do not themselves define the cure mechanism. Model selection should therefore begin with the scientific cure estimand and joint-distribution construction, followed by latency, observation-process, and estimation assumptions rather than by a model name or computational algorithm alone.

The taxonomy suggests that a competing-risk cure analysis should report, in a consistent order, the target cure estimand and its scope, the joint-distribution decomposition that generates cure, the event-time or CIF structure, the assumed cross-cause, cure--event, failure--censoring, and within-cluster relations, the treatment of masked causes, the observation scheme, and the estimation, uncertainty, and model-checking procedures. This information is especially important with interval censoring, masked causes, clustering, or substantial censoring, because superficially similar models may answer different scientific questions. Comparisons should distinguish structural assumptions from computational choices: an EM algorithm is an estimation device rather than a cure-model definition, and a neural-network M-step or a gradient-free optimizer does not by itself determine the decomposition or latency structure. Sensitivity analyses over cure definitions, tail and follow-up assumptions, dependence and censoring assumptions, latency families, and alternative estimators would make empirical conclusions more transparent and portable across applications.

The bone-marrow-transplant illustration shows how the five axes govern interpretation in practice. Defining relapse and death before relapse as first-event competing outcomes makes the reported global tail an event-free probability for that endpoint, not a cause-specific mortality or treatment-effect measure. The information criteria offered conflicting support for adding high-risk AML effects, and the group-specific global-tail bootstrap intervals overlapped. With limited sample size and follow-up, a restrictive defective Gompertz form, independent cause components, and independent censoring, the tail estimates are model-dependent extrapolations; the empirical survival and cumulative-incidence curves therefore remain essential complements.

Future work should develop a modular, versioned software framework that makes compatible choices of cure definition, joint-distribution decomposition, latency model, dependence structure, and estimation procedure explicit rather than implying that all components are freely interchangeable. Such a framework should support right- and interval-censored observations, clustered data, masked causes, zero-time failures, and latent-count mechanisms, while providing documented simulation benchmarks, unit tests, reproducible analysis scripts, uncertainty quantification, and an explicit software license. On the methodological side, further work is needed on identifiability under cause-specific cure and latent dependence, robust inference with dependent censoring and clustering, principled model checking for defective and latent-count formulations, and uncertainty assessment for highly flexible models or neural-network latent-count components. Common benchmark designs should record the target estimand and data-generating mechanism, so that methods are compared on more than predictive performance or numerical convergence.

In summary, the five-axis framework recasts a fragmented literature as a set of linked but distinct modeling choices rather than a universal ranking of methods. The illustrative application shows why cure estimates, fitted curves, and information criteria must be interpreted in relation to the chosen endpoint and model construction. Clear reporting, reproducible software, and common benchmarking standards are needed for meaningful comparison and use of competing-risk cure models.

\pagebreak

\bibliographystyle{jasa}
\bibliography{mybib}

\begin{thebibliography}{33}
\newcommand{\enquote}[1]{``#1''}
\expandafter\ifx\csname natexlab\endcsname\relax\def\natexlab#1{#1}\fi

\bibitem[\protect\citename{Basu and Tiwari, }2010]{basu_tiwari_2010}
Basu, S. and Tiwari, R.~C. (2010).
\newblock \enquote{Breast cancer survival, competing risks and mixture cure
  model: a Bayesian analysis.}
\newblock {\em Journal of the Royal Statistical Society Series A: Statistics in
  Society\/}, 173, 2, 307--329.

\bibitem[\protect\citename{Berkson and Gage, }1952]{berkson_gage_1952}
Berkson, J. and Gage, R.~P. (1952).
\newblock \enquote{Survival curve for cancer patients following treatment.}
\newblock {\em Journal of the American Statistical Association\/}, 47, 259,
  501--515.

\bibitem[\protect\citename{Boag, }1949]{boag_1949}
Boag, J.~W. (1949).
\newblock \enquote{Maximum likelihood estimates of the proportion of patients
  cured by cancer therapy.}
\newblock {\em Journal of the Royal Statistical Society. Series B
  (Methodological)\/}, 11, 1, 15--53.

\bibitem[\protect\citename{Chao, }1998]{chao_1998}
Chao, E.~C. (1998).
\newblock \enquote{Gibbs sampling for long-term survival data with competing
  risks.}
\newblock {\em Biometrics\/},  350--366.

\bibitem[\protect\citename{Chen et~al., }2020]{chen_et_al_2020}
Chen, C.-M., Shen, P.-s., Lin, C.-C., and Wu, C.-C. (2020).
\newblock \enquote{Semiparametric mixture cure model analysis with competing
  risks data: Application to vascular access thrombosis data.}
\newblock {\em Statistics in Medicine\/}, 39, 27, 4086--4099.

\bibitem[\protect\citename{Choi and Zhou, }2002]{choi_zhou_2002}
Choi, K. and Zhou, X. (2002).
\newblock \enquote{Large sample properties of mixture models with covariates
  for competing risks.}
\newblock {\em Journal of multivariate analysis\/}, 82, 2, 331--366.

\bibitem[\protect\citename{Choi et~al., }2015]{choi_et_al_2015}
Choi, S., Huang, X., and Cormier, J.~N. (2015).
\newblock \enquote{Efficient semiparametric mixture inferences on cure rate
  models for competing risks.}
\newblock {\em Canadian Journal of Statistics\/}, 43, 3, 420--435.

\bibitem[\protect\citename{Choi et~al., }2018]{choi_et_al_2018}
Choi, S., Zhu, L., and Huang, X. (2018).
\newblock \enquote{Semiparametric accelerated failure time cure rate mixture
  models with competing risks.}
\newblock {\em Statistics in medicine\/}, 37, 1, 48--59.

\bibitem[\protect\citename{Esmailian et~al., }2023]{esmailian_et_al_2023}
Esmailian, M., Azimi, R., Gallardo, D.~I., and Nasiri, P. (2023).
\newblock \enquote{New Cure Rate Survival Models Generated by Poisson
  Distribution and Different Regression Structures with Applications to Cancer
  Data Set.}
\newblock {\em Journal of Mathematics\/}, 2023, 1, 6292693.

\bibitem[\protect\citename{Farewell, }1982]{farewell_1982}
Farewell, V.~T. (1982).
\newblock \enquote{The use of mixture models for the analysis of survival data
  with long-term survivors.}
\newblock {\em Biometrics\/},  1041--1046.

\bibitem[\protect\citename{Ganguly et~al., }2026]{ganguly_et_al_2026}
Ganguly, A., Sultana, F., Kundu, D., and Pal, A. (2026).
\newblock \enquote{A Model Based on Mixture of Weibull Distributions for
  Depending Competing Risks Data in the Presence of Long-Term Survivors, and
  Its Application to Malignant Melanoma Cancer Data.}
\newblock {\em Statistics in Medicine\/}, 45, 6-7, e70466.

\bibitem[\protect\citename{Greenhouse and Wolfe, }1984]{greenhouse_wolfe_1984}
Greenhouse, J.~B. and Wolfe, R.~A. (1984).
\newblock \enquote{A competing risks derivation of a mixture model for the
  analysis of survival data.}
\newblock {\em Communications in statistics-Theory and Methods\/}, 13, 25,
  3133--3154.

\bibitem[\protect\citename{Kalbfleisch and Prentice,
  }2002]{kalbfleisch_prentice_2002}
Kalbfleisch, J.~D. and Prentice, R.~L. (2002).
\newblock {\em The statistical analysis of failure time data\/}.
\newblock John Wiley \& Sons.

\bibitem[\protect\citename{Kattumannil et~al., }2020]{kuttumannil_et_al_2020}
Kattumannil, S.~K. et~al. (2020).
\newblock \enquote{Semiparametric transformation model for competing risks data
  with cure fraction.}
\newblock {\em arXiv preprint arXiv:2007.02305\/}.

\bibitem[\protect\citename{Li et~al., }2007]{li_et_al_2007}
Li, Y., Tiwari, R.~C., and Guha, S. (2007).
\newblock \enquote{Mixture cure survival models with dependent censoring.}
\newblock {\em Journal of the Royal Statistical Society Series B: Statistical
  Methodology\/}, 69, 3, 285--306.

\bibitem[\protect\citename{Maller and Zhou, }1996]{maller_zhou_1996}
Maller, R. and Zhou, X. (1996).
\newblock {\em Survival Analysis with Long-Term Survivors\/}.
\newblock Wiley Series in Probability and Statistics. Wiley.

\bibitem[\protect\citename{Menger et~al., }2023]{menger_et_al_2023}
Menger, A., Sheikh, M.~T., and Chen, M.-H. (2023).
\newblock \enquote{Bayesian Modeling of Survival Data in the Presence of
  Competing Risks with Cure Fractions and Masked Causes.}
\newblock {\em Sankhya A\/},  1--29.

\bibitem[\protect\citename{Ng and McLachlan, }1998]{ng_mclachlan_1998}
Ng, S. and McLachlan, G. (1998).
\newblock \enquote{On modifications to the long-term survival mixture model in
  the presence of competing risks.}
\newblock {\em Journal of Statistical Computation and Simulation\/}, 61, 1-2,
  77--96.

\bibitem[\protect\citename{Nicolaie et~al., }2019]{nicolaie_et_al_2019}
Nicolaie, M.~A., Taylor, J.~M., and Legrand, C. (2019).
\newblock \enquote{Vertical modeling: analysis of competing risks data with a
  cure fraction.}
\newblock {\em Lifetime data analysis\/}, 25, 1--25.

\bibitem[\protect\citename{Othus et~al., }2009]{othus_et_al_2009}
Othus, M., Li, Y., and Tiwari, R.~C. (2009).
\newblock \enquote{A class of semiparametric mixture cure survival models with
  dependent censoring.}
\newblock {\em Journal of the American Statistical Association\/}, 104, 487,
  1241--1250.

\bibitem[\protect\citename{Pal and Roy, }2026]{pal_roy_2026}
Pal, S. and Roy, S. (2026).
\newblock \enquote{A new estimation algorithm for destructive cure model:
  illustration with exponentially weighted Poisson competing risks.}
\newblock {\em Communications in Statistics-Simulation and Computation\/},
  1--15.

\bibitem[\protect\citename{Putter et~al., }2007]{putter_et_al_2007}
Putter, H., Fiocco, M., and Geskus, R.~B. (2007).
\newblock \enquote{Tutorial in biostatistics: competing risks and multi-state
  models.}
\newblock {\em Statistics in medicine\/}, 26, 11, 2389--2430.

\bibitem[\protect\citename{Rejani and Sankaran, }2020]{rejani_sankaran_2020}
Rejani, P. and Sankaran, P. (2020).
\newblock \enquote{Modeling and analysis of proportional hazards competing
  risks cure rate model.}
\newblock {\em Journal of the Indian Society for Probability and Statistics\/},
  21, 175--185.

\bibitem[\protect\citename{Silpa et~al., }2026]{silpa_et_al_2024}
Silpa, K., Sreedevi, E., and Sankaran, P. (2026).
\newblock \enquote{Defective regression models for cure rate data with
  competing risks.}
\newblock {\em Journal of Biopharmaceutical Statistics\/}, 36, 3, 381--397.

\bibitem[\protect\citename{Silpa et~al., }2024]{silpa_et_al_2024a}
Silpa, K., Sreedevi~E, P., and Sankaran, P. (2024).
\newblock \enquote{Regression Analysis of Cure Rate Models with Competing Risks
  Subjected to Interval Censoring.}
\newblock {\em arXiv preprint arXiv:2412.04803\/}.

\bibitem[\protect\citename{Sreedevi et~al., }2026]{sreedevi_et_al_2026}
Sreedevi, E., Silpa, K., and Sankaran, P. (2026).
\newblock \enquote{Gompertz regression model for zero-inflated cure rate data
  with competing risks: an application to cutaneous melanoma data.}
\newblock {\em Japanese Journal of Statistics and Data Science\/},  1--24.

\bibitem[\protect\citename{Sy and Taylor, }2000]{sy_taylor_2000}
Sy, J.~P. and Taylor, J.~M. (2000).
\newblock \enquote{Estimation in a Cox proportional hazards cure model.}
\newblock {\em Biometrics\/}, 56, 1, 227--236.

\bibitem[\protect\citename{Teh et~al., }2025]{teh_et_al_2025}
Teh, L.~R., Cancho, V.~G., and Rodrigues, J. (2025).
\newblock \enquote{Modeling of long-term survival data with unobserved
  dispersion via neural network: LR Teh et al.}
\newblock {\em Computational Statistics\/}, 40, 8, 4115--4137.

\bibitem[\protect\citename{Wang et~al., }2026{\natexlab{a}}]{wang_et_al_2026a}
Wang, Y., Li, J., Zhou, Q., and Wang, W. (2026{\natexlab{a}}).
\newblock \enquote{A unified framework for complex survival data: Accounting
  for clustering, cure fractions, and competing risks.}
\newblock {\em BMC Medical Research Methodology\/}.

\bibitem[\protect\citename{Wang et~al., }2020{\natexlab{a}}]{wang_et_al_2020a}
Wang, Y., Tang, Y., and Zhang, J. (2020{\natexlab{a}}).
\newblock \enquote{Bayesian approach for proportional hazards mixture cure
  model allowing non-curable competing risk.}
\newblock {\em Journal of Statistical Computation and Simulation\/}, 90, 4,
  638--656.

\bibitem[\protect\citename{Wang et~al., }2021]{wang_et_al_2021}
Wang, Y., Zhang, J., Cai, C., Lu, W., and Tang, Y. (2021).
\newblock \enquote{Semiparametric estimation for proportional hazards mixture
  cure model allowing non-curable competing risk.}
\newblock {\em Journal of Statistical Planning and Inference\/}, 211, 171--189.

\bibitem[\protect\citename{Wang et~al., }2020{\natexlab{b}}]{wang_et_al_2020b}
Wang, Y., Zhang, J., and Tang, Y. (2020{\natexlab{b}}).
\newblock \enquote{Semiparametric estimation for accelerated failure time
  mixture cure model allowing non-curable competing risk.}
\newblock {\em Statistical Theory and Related Fields\/}, 4, 1, 97--108.

\bibitem[\protect\citename{Wang et~al., }2026{\natexlab{b}}]{wang_et_al_2026b}
Wang, Y., Zhou, B., Fu, C., and Xu, X. (2026{\natexlab{b}}).
\newblock \enquote{Modeling cure fractions in interval-censored competing risks
  data: An AFT mixture approach.}
\newblock {\em Electronic Research Archive\/}, 34, 9, 6573--6613.

\end{thebibliography}

\end{document}